\documentclass[11pt]{article}
\usepackage[intlimits]{amsmath}
\usepackage{lineno}
\usepackage{amsfonts}
\usepackage{amssymb}
\usepackage{fancyhdr}
\usepackage{epsfig}
\usepackage{rotating}
\usepackage{lscape}
\usepackage{graphicx}
\newcommand {\rusl}   {\ensuremath{R_{u/sl}}}

\def\B      {\ensuremath{B}\hbox{ }}
\def\Bp      {\ensuremath{B^{+}}\hbox{ }}
\newcommand {\rhop} {\ensuremath{\rho^+}\hbox{ }}
\newcommand {\rhoz} {\ensuremath{\rho^0}\hbox{ }}
\newcommand {\rhozv} {\mbox{``\ensuremath{\rho^0}''}\hbox{ }}
\newcommand {\az} {\ensuremath{a^0_{0}}\hbox{ }}
\newcommand {\azp} {\ensuremath{a^+_{0}}\hbox{ }}
\newcommand {\bpiz} {\ensuremath{\Bm \rightarrow \piz \ell \bar{\nu}}\hbox{ }}
\newcommand {\bpi} {\ensuremath{\Bzb \rightarrow \pip \ell \bar{\nu}}\hbox{ }}
\newcommand {\bomega} {\ensuremath{\Bm \rightarrow \omega \ell \bar{\nu}}\hbox{ }}
\newcommand {\brhop} {\ensuremath{\Bzb \rightarrow \rho^+ \ell \bar{\nu}}\hbox{ }}
\newcommand {\brhoz} {\ensuremath{\Bm \rightarrow \rho^0 \ell \bar{\nu}}\hbox{ }}
\newcommand {\bet} {\ensuremath{\Bm \rightarrow \eta \ell \bar{\nu}}\hbox{ }}
\newcommand {\betp} {\ensuremath{\Bm \rightarrow \etapr \ell \bar{\nu}}\hbox{ }}
\newcommand {\baz} {\ensuremath{\Bm \rightarrow a^0_{0} \ell \bar{\nu}}\hbox{ }}
\newcommand {\bazp} {\ensuremath{\Bzb \rightarrow a^+_{0} \ell \bar{\nu}}\hbox{ }}
\newcommand {\brhozv} {\ensuremath{\Bm \rightarrow  \rhozv \ell \bar{\nu}}\hbox{ }}
\newcommand {\bulnu}{\ensuremath{\b \rightarrow \u \ell \bar{\nu}}\hbox{ }}
\newcommand {\bclnu}{\ensuremath{\b \rightarrow \c \ell \bar{\nu}}\hbox{ }}

\newcommand {\Bxclnu}{\ensuremath{\Bb \rightarrow X_c \ell \bar{\nu}}}
\newcommand {\Bxulnu}{\ensuremath{\Bb \rightarrow X_u \ell \bar{\nu}}}
\newcommand {\Bxlnu}{\ensuremath{\Bb \rightarrow X \ell \bar{\nu}}}
\newcommand {\Bnxlnu}{\ensuremath{\Bzb \rightarrow X \ell \bar{\nu}}}
\newcommand {\Bpxlnu}{\ensuremath{\B^{-} \rightarrow X \ell \bar{\nu}}}

\newcommand {\mX}{\ensuremath{m_{X}}}
\newcommand {\mmiss}{\ensuremath{m_{miss}^2}}

\newcommand {\breco}{\ensuremath{B_{reco}}}

\newcommand{\beq}{\begin{equation}}
\newcommand{\beqa}{\begin{eqnarray}}
\newcommand{\beqn}{\begin{eqnarray}}
\newcommand{\eeq}{\end{equation}}
\newcommand{\eeqa}{\end{eqnarray}}
\newcommand{\eeqn}{\end{eqnarray}}
\makeatletter
\def\slash#1{{\mathpalette\c@ncel{#1}}} 
\makeatother

\newcommand{\upr}[1]{\mathrm{#1}}

\newcommand{\btoulnu}{$b\rightarrow u \ell \bar{\nu}$}

\newcommand {\mx}     {\ensuremath{M_{X}}}
\newcommand {\Q}      {\ensuremath{q^{2}}}
\newcommand{\xmeas}{$\overrightarrow{x}^\upr{meas}$}
\newcommand{\xtru}{$\overrightarrow{x}^\upr{tru}$}
\newcommand{\xtau}{$\overrightarrow{x}^{{\rm unf}}$}

\newcommand{\Ctau}{$\widehat{C}^{\rm unf}$}

\newcommand{\unit}[1]{\,\mathrm{#1}}
\newcommand{\bs}{ {\bar{s}} }

\newcommand{\mbsf} {\ensuremath{m_b^{SF}}}
\newcommand{\lonesf} {\ensuremath{\lambda_1^{SF}}}
\newcommand{\lbarsf} {\ensuremath{\bar{\Lambda}^{SF}}}
\newcommand {\brecoil}{\ensuremath{B_{recoil}}\hbox{ }}

\newcommand{\gevccsq}{\ensuremath{{\mathrm{\,Ge\kern -0.1em V^2\!/}c^4}}\xspace}

\newcommand{\BABARPubYear}    {04}

\newcommand{\BABARConfNumber} {11}
\newcommand{\SLACPubNumber} {10651}
\newcommand{\LANLNumber} {0408068}

\input pubboard/babarsym

\newcounter{TODO}

\long\def\inst#1{\par\nobreak\kern 4pt\nobreak
    {\it #1}\par\vskip 10pt plus 3pt minus 3pt}

\begin{document}

{\pagestyle{empty}

\begin{flushright}
\babar-CONF-\BABARPubYear/\BABARConfNumber \\
SLAC-PUB-\SLACPubNumber \\
hep-ex/\LANLNumber \\
\end{flushright}

\par\vskip 1.2cm
\begin{center}
\Large \bf Study of \boldmath\btoulnu\ Decays on the Recoil of Fully
Reconstructed B Mesons  and Determination of \boldmath \Vub
\end{center}
\bigskip

\begin{center}
\large The \babar\ Collaboration\\
\mbox{ }\\
\today
\end{center}
\bigskip \bigskip

\begin{center}
\large \bf Abstract
\end{center}
Based on 88 million \FourS\to\BB decays collected by the \babar\ experiment at the PEP-II~asymmetric-energy $B$
factory at  SLAC,
 we report preliminary results of four analyses which investigate semileptonic charmless $B$ decays,
 \Bxulnu. Deeper understanding of all aspects of these decays will improve the determination of the 
Cabibbo-Kobayashi-Maskawa matrix element \Vub. In events in which one $B$ meson decay to a hadronic 
final state is fully reconstructed, the semileptonic decay of the second $B$ meson is identified 
by the detection of a charged lepton. 
By measuring the spectrum of the invariant mass of the hadronic system $X_u$ (\mx), we derive the
branching fraction
  $\BR(\Bxulnu)=(2.53\pm 0.29(\rm stat.)\pm 0.26(\rm
  sys.)^{+0.69}_{-0.41} (\rm theo.))\times 10^{-3}$.
The two-dimensional distribution of \mx\ and \Q, the squared lepton-neutrino invariant mass, 
is used to derive the partial branching fraction for $\mx<1.7\gevcc,\Q>8\gevccsq$ to be 
$\Delta\BR(\Bxulnu )=(0.88\pm 0.14(\rm stat.)\pm 0.13(\rm sys.)\pm 0.02(\rm
theo.))\times 10^{-3}$. 
From these two measurements we can extract
$\Vub=(4.77\pm 0.28(\rm stat.)\pm 0.28 (\rm sys.)^{+0.69}_{-0.45}(\rm theo.))\times 10^{-3}$ and 
$\Vub=(4.92\pm 0.39(\rm stat.)\pm 0.36 (\rm sys.)\pm 0.46(\rm theo.))\times 10^{-3}$, respectively.
We use the same sample to extract  the true \mx\ distribution for \Bxulnu\ events, with the goal of  
comparing it with theoretical models. 
We  also identify several exclusive charmless semileptonic $B$ decays, 
and measure the branching fractions 
$ \BR(\Bbar\rightarrow \pi\ell\nub) = (1.08 \pm 0.28(\rm stat.) \pm 0.16(\rm sys.))\times 10^{-4}$ and  
$\BR(\Bbar\rightarrow\rho\ell\nub) = (2.57 \pm 0.52(\rm stat.) \pm 0.59(\rm sys.))\times 10^{-4}$ using isospin and quark model constraints. 
We also set limits on  \BR(\bet), ~\BR(\betp), 
~$\BR(\baz)\BR(\az \rightarrow \eta \piz)$, ~and $\BR(\bazp)\BR(\azp \rightarrow \eta \pip)$.

\vfill
\begin{center}

Submitted to the 32$^{\rm nd}$ International Conference on High-Energy Physics, ICHEP 04,\\
16 August---22 August 2004, Beijing, China

\end{center}

\vspace{1.0cm}
\begin{center}
{\em Stanford Linear Accelerator Center, Stanford University, 
Stanford, CA 94309} \\ \vspace{0.1cm}\hrule\vspace{0.1cm}
Work supported in part by Department of Energy contract DE-AC03-76SF00515.
\end{center}

\newpage
} 

\begin{center}
\small

The \babar\ Collaboration,
\bigskip

%
B.~Aubert,
R.~Barate,
D.~Boutigny,
F.~Couderc,
J.-M.~Gaillard,
A.~Hicheur,
Y.~Karyotakis,
J.~P.~Lees,
V.~Tisserand,
A.~Zghiche
\inst{Laboratoire de Physique des Particules, F-74941 Annecy-le-Vieux, France }
% ...

\end{center}\newpage

\section{Introduction}
The principal physics goal of the \babar\ experiment is to establish
\CP violation in $B$ mesons and to test whether the observed effects
are consistent with the predictions of the Standard Model (SM). 
\CP violating effects result in the SM from an irreducible phase in the 
Cabibbo-Kobayashi-Maskawa (CKM) matrix which describes
the couplings of the charged weak current to quarks. An improved
determination of the absolute value of the matrix element
\Vub, the coupling strength of the $b$ quark to the $u$ quark, 
will contribute critically to tests of the consistency of the 
angles of the unitarity triangle of the CKM matrix. 

The extraction of \Vub\ is a challenge both for theory and
experiment.  Experimentally, the main problem is the separation of
$b\rightarrow u \ell\nu$ decays from the more abundant $b\rightarrow c
\ell \nu$ decays.  Selection criteria applied to achieve this
separation generally make the theoretical extrapolation to the full
decay rate more difficult.
Theoretically, inclusive semileptonic rates can be calculated
reliably at the parton level. However, the dynamics of $B$ meson decays depend
on the $b$ quark mass and its motion inside the 
meson. Calculations of the decay rate rely on
operator product expansions (OPE) in inverse powers of the $b$ quark
mass. These depend on the choice of renormalization scale and
include non-perturbative contributions, resummed into the so-called shape function (SF), which introduce model uncertainties.
Exclusive branching fractions can also  be related to \Vub, although with large  errors. In addition, 
the as yet poorly known dynamics of the decays introduce large uncertainties in the determination of the efficiencies of the selection criteria.
To reduce these uncertainties we apply as loose selection criteria as possible to
exclusive \Bxulnu\ decays\footnote{Unless otherwise specified, charge conjugation is always implied 
throughout this paper.}. 

\begin{figure}[htb]
 \begin{centering} 
 \epsfig{file=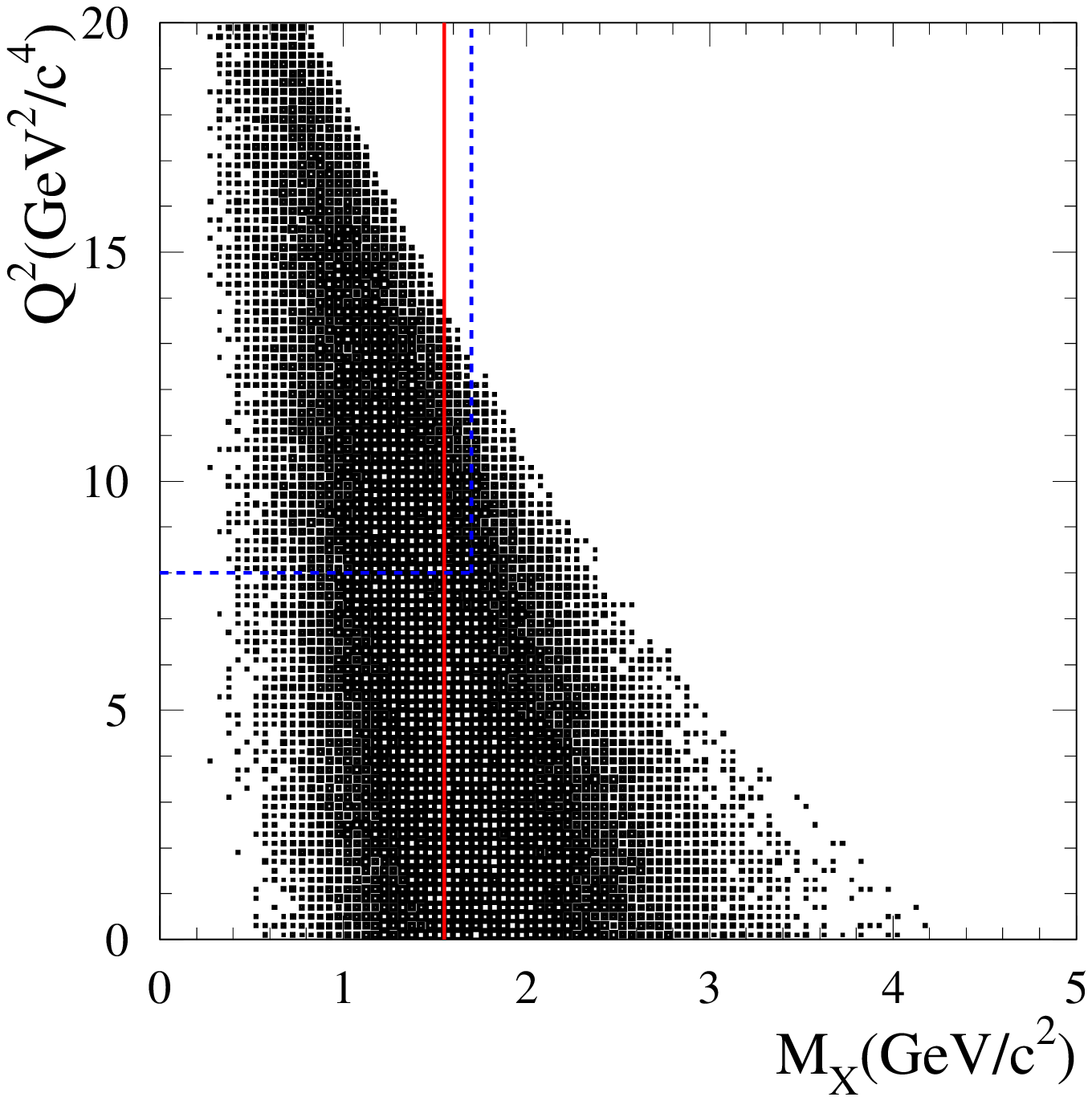,width=0.48\textwidth}
 \epsfig{file=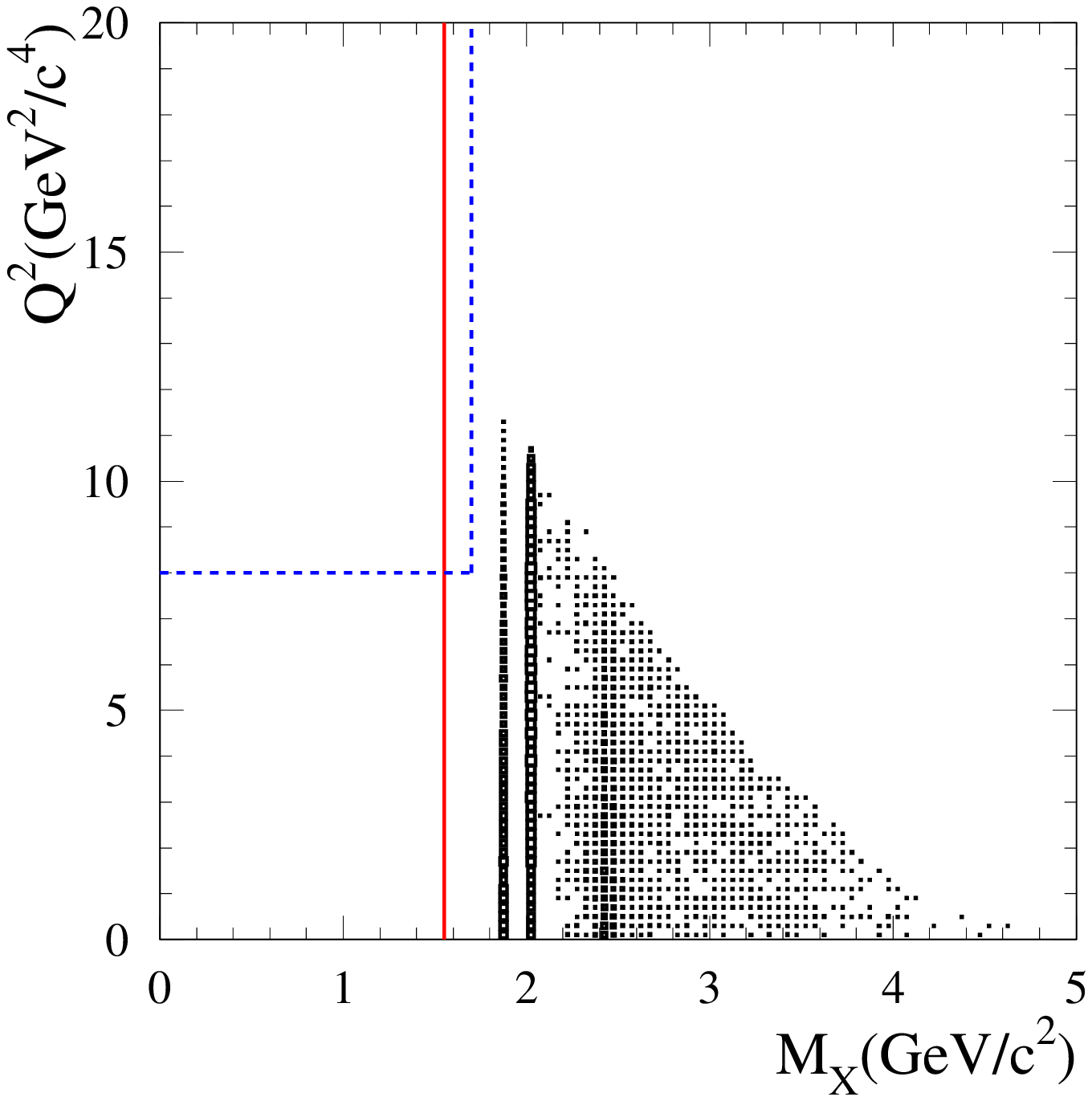,width=0.48\textwidth}
\caption{Distribution on MonteCarlo simulated events of the squared invariant mass of the leptons (\Q)  and  the 
invariant mass of the hadronic recoil system (\mx) in semileptonic \bulnu\ (left) and 
 \bclnu\ decays (right). The model utilized is the non-resonant model described in Sec.~\ref{sec:theosys}. The full (dashed) line indicates
 the phase-space region selected by the \mx (\mx-\Q)  analysis.}
\label{fig:mxq2gen}
\end{centering}
\end{figure}

At the \FourS\ resonance, events with a reconstructed $B$ decay to hadronic final states are an optimal 
environment for the study of semileptonic decays of the second $B$ meson. 
The relatively small backgrounds allow for loose selection criteria and reduce the uncertainty in the 
extrapolation to the full decay rate. Moreover, the fact that the $B$ momentum is known 
allows us to isolate the signal in several regions of phase space and 
perform different measurements with relatively uncorrelated theoretical and experimental uncertainties. 
The kinematic variables considered in this paper are \mx, the invariant
mass  of the hadrons $X$, and \Q, the squared invariant mass of the two leptons.
The main background from \Bxclnu\ decays is located at high values
of \mx\ and low values of \Q, while the \Bxulnu\ signal events extends to
low \mx\ and high \Q\ values (see Fig~\ref{fig:mxq2gen}).

In this paper we present several preliminary results that extend the \Vub\ analysis 
published by \babar\ in Ref.~\cite{Aubert:2003zw}. These new studies are based on the same analysis strategy 
and data set. They are motivated by the recent discussion 
on the theoretical uncertainties to be assigned to the measurement of \Vub, 
see for instance~\cite{Bauer:2003pi,Bosch:2004bt,Gibbons:2004dg} 
and by the possibility of studying exclusive charmless decays with high purity. 
This paper presents four separate results:
\begin{itemize}
\item A \Vub measurement is performed  utilizing the \mx\ spectrum, as in Ref.~\cite{Aubert:2003zw},
following the approach from  De~Fazio and Neubert~\cite{DeFazio:1999sv}
referred to as ``DFN'' in the following.
An updated estimate of the SF parameters is used 
for the 
extrapolation to the full phase space. This analysis will be referred to as the ``\mx'' analysis.
\item A determination of the true \mx\ spectrum with
  detector effects unfolded (``unfolding'' analysis). This spectrum 
allows for a direct comparison with theoretical 
models and with more statistics will be able to constrain the SF parameters. We determine the cumulative \mx\ distribution, as well as 
the first and second mass moment of the \mx\ distribution.
\item A \Vub measurement is performed by utilizing the two-dimensional
  distribution of \mx\ and \Q\ and by following  the approach of Bauer, Ligeti and 
Luke~\cite{Bauer:2001yb}, referred to as ``BLL'' in the following. This analysis, which will be referred to as 
the ``\mx-\Q'' analysis, requires the determination of the partial branching fraction
in limited regions of the phase space.
\item A measurement of branching fractions for exclusive charmless semileptonic $B$ decays.  
In this case the semileptonic decay is identified by a charged lepton, and the hadronic state is exclusively reconstructed.
We analyze the following decay modes:
\bpi, \bpiz, \brhop, \brhoz, \bomega, \bet, \betp, \baz and \bazp.

\end{itemize}

A review of the previous measurements of \Vub\ is given in Ref.~\cite{Gibbons:2004dg}. The measurements reported 
here are either based on novel techniques or have smaller uncertainties than the existing ones.

This paper is organized as follows:
section~\ref{sec:datasample} describes the detector, the data sample and the Monte Carlo simulation including 
a description of the theoretical model on which our efficiency calculations are based.
Section~\ref{sec:strategy} 
describes the aspects common to all analyses presented here.
Section~\ref{sec:mx} presents a new interpretation of the published \mx\
analysis that takes the latest theoretical developments into account. 
Section~\ref{sec:unfold} discusses the results of the \mx\ spectrum unfolding (on the whole
\Q\ range). 
Section~\ref{sec:mxq2} shows the details and the results of the
\mx-\Q\ analysis, while Section~\ref{sec:belle} shows the results for an
alternative set of SF parameters. 
Finally in Section~\ref{sec:excl} we present the measurement of exclusive \Bxulnu\ branching fractions.

\section{Data Sample and Simulation}
\label{sec:datasample}
The data used in this paper were recorded with the \babar\ 
detector~\cite{Aubert:2001tu}
at the \pep2\ collider in the period October 1999--September 2002.
The total integrated luminosity of the data set is 
81.9 \invfb\ collected on the \FourS\ resonance.  The corresponding number
of produced \BB\ pairs is 88 million. 
We use Monte Carlo (MC) simulations of the \babar\ detector based on
\geant~\cite{geant} to optimize selection criteria and to
determine signal efficiencies and background distributions.

\subsection{Simulation of \Bxulnu\ decays}
\label{sec:theosys}
Charmless semileptonic \Bxulnu\ decays  are simulated as a combination  of  both 
three-body decays to narrow resonances, $X_u = \pi,  \eta,  \rho,  \omega,   \eta^\prime$,  
and  decays to non-resonant hadronic final states $X_u$.
\begin{figure}[htb]
 \begin{centering} 
\centerline{\epsfig{figure=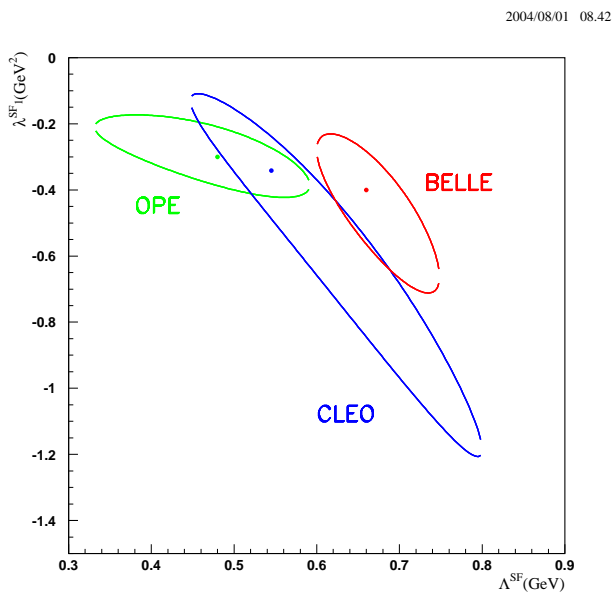,width=0.5\linewidth}}
\caption{$\Delta\chi^2=1$ contours for the fit to $b\to c\ell\nu$ moments in Ref.~\cite{Cronin-Hennessy:2001fk}(green) and the $b\to s\gamma$ photon energy spectra in Belle~\cite{tadao} (red) and CLEO~\cite{Gibbons:2004dg} (blue). Dots represent the best $\chi^2$ points.}
\label{fig:ellipses}
\end{centering}
\end{figure} 

The simulation of the inclusive charmless semileptonic $\B$ decays into hadronic 
states with masses larger than $2m_{\pi}$  is based on a prescription 
by De~Fazio and Neubert~\cite{DeFazio:1999sv} (DFN), which calculates the
triple differential decay rate,  $d^3\Gamma\,/\,dq^2\,dE_{\ell}\,ds_H$
($s_H=\mX^2$), up to ${\cal O}(\alpha_{\rm s})$  corrections.
The motion of the $\b$ quark inside the $\B$ meson 
is incorporated in the DFN formalism by convolving  the parton-level triple
differential decay rate with a non-perturbative shape function (SF). 
The SF describes the distribution of the momentum $k_+$  of the $\b$ quark inside the $\B$ meson.  
The two free parameters of the SF are 
$\lbarsf$ and $\lonesf$. The first relates the $B$ meson mass, $m_B$, to the
$b$ quark mass, $\mbsf = m_B - \lbarsf$, and
$-\lonesf$ is the average momentum squared of the ${\b}$ quark in
the ${\B}$ meson.
The SF parameterization used in the generator is of the form  
 \begin{equation}
  F(k_+) = N (1-x)^a e^{(1+a)x},
 \label{eq:fermi_motion}
 \end{equation}
where $x = \frac{k_+}{\lbarsf} \le 1 $ and $a = -3 (\lbarsf)^{2}/{\lonesf}-1$. 
The original DFN paper~\cite{DeFazio:1999sv} suggested that the SF parameters could be related to 
the  operator product expansion  parameters  by $m_B - \lbarsf=m_b$ and $\lonesf=-\mu_\pi^2$. 
Since there is currently no consensus on these relationships (see for instance Refs.~\cite{Bauer:2003pi,Bosch:2004bt}) we  choose to
 use the values of \lbarsf\ and \lonesf extracted from the $b\to s\gamma$ spectrum by CLEO~\cite{Gibbons:2004dg}. 
Fitting the spectrum to 
 simulated samples with different SF parameters allows us to create a
 $\Delta\chi^2=1$ contour, shown in Figure~\ref{fig:ellipses}, which
 we use to estimate theoretical uncertainties. 
The point with minimum $\chi^2$ corresponds to $\lbarsf = 0.545 \gev/c^2$ and
$\lonesf=-0.342\gev^2/c^4$. 
\begin{figure}
\begin{centering}
\epsfig{file=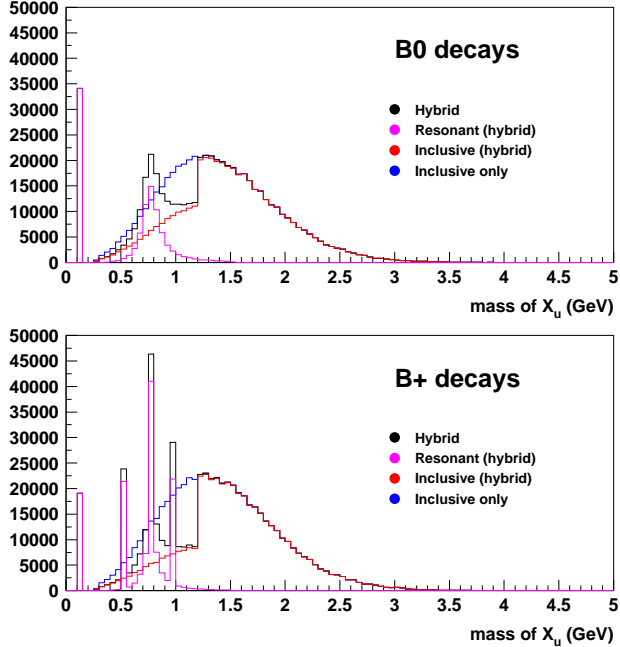,width=0.60\textwidth}

\caption{ Model for the simulation of  \bulnu\ decays for neutral (top)
  and charged (bottom) $B$ mesons: resonant decays (purple - peaky) are 
combined with  weighted non-resonant (``inclusive'') simulated events
(red - smooth) to form the 
``hybrid'' model (black). The inclusive sample is also shown before applying the weights (blue).
}
\label{fig:mxhadgen}
\end{centering}
\end{figure}

In the simulation 
the hadron system $X_u$ is produced with a non-resonant
and continuous invariant mass spectrum according to the DFN model.
A reweighting  of the Fermi  motion distribution is 
used to  obtain distributions for different values of  \lbarsf\ and
\lonesf.  Finally, 
the fragmentation of the $X_u$ system into final state hadrons 
is performed by JETSET~\cite{Sjostrand:1994yb}.
The exclusive charmless semileptonic decays are simulated using 
the ISGW2  model~\cite{isgw2}.
The resonant and non-resonant components are combined such that the total branching fraction is 
consistent with the measured value~\cite{Aubert:2003zw} and that the integrated spectrum agrees 
with the prediction of Ref.~\cite{DeFazio:1999sv}. 
%
The resulting \mX\ distributions for charged and neutral $B$ mesons are shown in 
Fig.~\ref{fig:mxhadgen}. All branching fractions and theory parameters involved in this reweighting are varied within their 
errors in the evaluation of the associated uncertainty.

While we were writing this paper we became aware of a preliminary  interpretation of the $b\to s\gamma$ energy spectrum measured
 by Belle~\cite{tadao}. 
The resulting $\Delta\chi^2=1$ contour is shown in Fig.~\ref{fig:ellipses}: it is consistent with the CLEO one and 
has smaller  uncertainties. A visible shift of the central value of the measured quantities  is expected though because the 
best fit result from Belle is outside the 1$\sigma$ CLEO contour.
Since there was not enough
time to combine the two results, from Belle and CLEO, we prefer to report the results with the CLEO ellipse as primary results
 and quote the results with the Belle ellipse as a reference for future developments. 

\subsection{Theoretical Acceptance Estimates}
The \mx-\Q\ analysis results in the partial branching fraction for charmless semileptonic 
decays in selected phase space regions. To translate it into a measurement of the total 
branching fraction, and therefore \Vub, we need the fraction of events inside the measurement region 
(referred to as ``acceptance'' in the rest of the paper) as an external input. 
We choose the results of Bauer, Ligeti and Luke \cite{Bauer:2001yb} for the acceptance corrections. These authors 
perform an operator-product-expansion-based calculation which includes ${\cal{O}}(\alpha_s^2)$ and ${\cal{O}}(1/m_b^2)$ 
corrections in a region of phase space where the non-perturbative effects due to the shape function are small. 
Their results show that the theoretical uncertainties due to the extrapolation to the full phase space can 
be significantly reduced with respect to a measurement based on a single \mx\ cut by
 moving the \mx\ cut to the highest practical value allowed by the charm semileptonic background, and by 
 reducing the \Q\ cut as low as possible.
As a cross-check, acceptances computed with the DFN model presented in the 
previous section are also used in the \mx-\Q\ analysis. 

\subsection{Simulation of Background Processes}
In order to estimate the shape of the background distributions we make use of simulations of 
$\epem\to\FourS\to\BB$ with the $B$ mesons
decaying inclusively. 
The most relevant backgrounds are due to  \Bxclnu\ events. 
The simulation of these processes uses an HQET parametrization of form factors for $\Bb\to
D^{*}\ell\nub$~\cite{Duboscq:1996mv}, and models  for $\Bb\to
D^{(*)} \pi \ell\nub$~\cite{Goity:1995xn}, and  $\Bb\to D
\ell\nub,D^{**}\ell\nub $~\cite{isgw2}.


\section{Common Analysis Aspects}
\label{sec:strategy}
The event selection and reconstruction and the measurements of branching fractions 
follow closely the strategy described in Ref.~\cite{Aubert:2003zw} and represents the common 
base for the analyses presented here. However, each analysis may differ in some details. 
This is particularly true for the measurement of exclusive branching fractions. 
This section describes aspects in common to all studies. Details specific to each 
analysis are provided in the following sections. 

\subsection{Event Reconstruction and Selection}

In this paper  we study the recoil of   fully reconstructed $B$ 
in hadronic decay modes (\breco), 
which is a moderately pure 
sample with known flavor and four-momentum.
We select \breco\ decays of the type  $B \rightarrow \Db  Y$,
where $D$ refers to a charm  meson, and $Y$ represents a collection of
hadrons with  a total  charge of $\pm  1$, composed  of $n_1\pi^{\pm}+
n_2K^{\pm}+ n_3\KS +  n_4\piz$, where $n_1 + n_2 < 6$,  $n_3 < 3$, and
$n_4 <  3$.  Using $D^-$  and $D^{*-}$ ($\Dzb$ and  $\Dstarzb$) as
seeds for  $B^0$ ($B^+$) decays,  we reconstruct about  1000 different
decay chains. Overall, we correctly reconstruct one $B$ candidate in 0.3\%  (0.5\%) of the 
\BzBzb\ (\BpBm) events.
The kinematic consistency of a $B_{reco}$ candidate with a $B$ meson
decay is checked using two variables, the beam-energy-substituted mass
$\mes = \sqrt{s/4 - \vec{p}^{\,2}_B}$
 and the energy difference, 
$\Delta E  = E_B  - \sqrt{s}/2$. Here  $\sqrt{s}$ refers to  the total
energy in the \FourS center of  mass frame, and $\overrightarrow{p_B}$ and $E_B$ denote
the momentum and energy of the $B_{reco}$ candidate in the same frame. 
For signal events the \mes\ distribution peaks at the $B$ meson mass, while
$\Delta E $ is consistent with zero. We  require $\Delta E  = 0$ within  approximately three
standard deviations.

A semileptonic decay of the other $B$ meson (\brecoil) is identified by the
presence of a charged lepton with momentum in the \brecoil\ rest frame ($p^*$) higher than 1\gevc\ . 
In addition, the detection of missing
energy and momentum in the event is taken as evidence for the presence
of a neutrino. 
The hadronic system $X$  is reconstructed from charged
tracks  and energy depositions
in the calorimeter that are not associated with
the \breco\ candidate or the identified lepton.  Care is taken to eliminate 
fake charged tracks, as well as low-energy beam-generated photons and energy depositions in the calorimeter from charged and
neutral hadrons.  The
neutrino four-momentum $p_{\nu}$ is estimated from the
missing momentum four-vector $p_{miss} = p_{\Upsilon(4S)}-p_{\breco} -p_X-p_\ell$, 
where all momenta are measured in the laboratory frame, and
$p_{\Upsilon(4S)}$ refers to the  \FourS\ meson. 

Undetected particles and measurement uncertainties affect the determination 
of the four-momenta of the  $X$ system and neutrino, 
 and lead to a large leakage of \Bxclnu\ background from the high \mx\
 into the low \mx\ region.
To improve the resolution on these four-momenta this analysis
exploits the well-known kinematics of the $e^+e^-$\to\FourS\to\BB\ process 
and performs a two-constraint kinematic fit to the whole event.  

In the sample of reconstructed $B$ decays two backgrounds need
to be considered: the combinatorial background from \BB\ and continuum events, 
due to random association of tracks and neutral clusters, which does not peak
in \mes, and the \BB\ background whose \mes\ distribution has the same shape as the signal.
After applying all selection criteria, the remaining combinatorial
background is subtracted by performing an unbinned likelihood fit
to the \mes\ distribution. In this fit, the combinatorial
background originating from $e^+e^-\to q\bar{q}$ ($q=u,d,s,c$) continuum
and \BB\ events is described by an empirical threshold function~\cite{Albrecht:1993pu}, and 
the signal is described by a modified Gaussian~\cite{cry} peaked at the $B$ meson mass. 

To reject the background in the sample of semileptonic decays we require exactly one
charged lepton with $p^*>$1\gevc, 
a total event charge of zero, and a missing mass consistent with zero
($\mmiss < 0.5 \gev^2/c^4$).  These criteria partly suppress the dominant
\Bxclnu\ decays, many of which contain an additional neutrino or
an undetected $\KL$ meson.  We explicitly veto
the $\Bzb\to\Dstarp\ell^-\overline{\nu}$ background by searching candidates for such a decay with a partial reconstruction technique, that is 
only identifying
the $\pi^+_s$ from the $\Dstarp\to \Dz\pi_s^+$ decay and the
lepton: since the momentum of the $\pi^+_s$ is almost collinear with the
\Dstarp\ momentum in the laboratory frame, we can approximate the
energy of the \Dstarp\ as $E_{\Dstarp}
\simeq m_{\Dstarp} \cdot E_{\pi_s} /145 \mevcc$ and estimate the neutrino mass as
  $  m_{\nu}^2  =  (p_B   -  p_{\Dstarp}  -  p_{\ell})^2 $. Events with $  m_{\nu}^2 >- 3  \gev^2/c^4$
are likely to be background events and are rejected.
Finally, we  veto events  with charged or  neutral kaons  in the
recoil  \Bb\ to reduce  the peaking background  from \Bxclnu\  decays.  Charged
kaons are identified \cite{Aubert:2001tu}
with an efficiency varying between 60\% at the highest and almost 100\% at the
lowest  momenta.  The  pion   misidentification rate is  about  2\%.   The
$\KS\to\pi^+\pi^-$  decays  are reconstructed  with  an efficiency  of
$80\%$ from pairs of oppositely  charged tracks with an invariant mass
between 486  and   510  \mevcc.

\subsection{Measurement of Charmless Semileptonic Branching Ratios}

\begin{figure}[htb]
 \begin{centering}
\epsfig{file=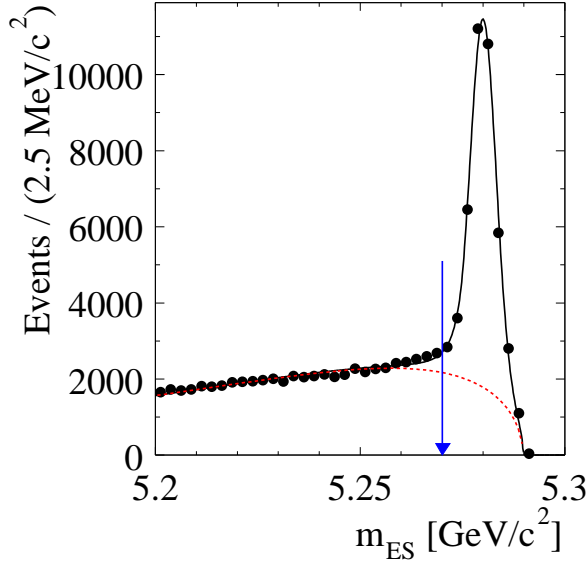,width=0.5\linewidth}
\caption{ Fit to the $\mes$ distribution for the sample with high momentum lepton. 
\label{fig:mxhadfit}}
 \end{centering}
\end{figure}

To  reduce the systematic uncertainties in the 
derivation of branching fractions,  the observed
number  of  signal events,  corrected   for  peaking background  and  efficiency,  is
normalized to the total number of semileptonic decays \Bxlnu\
 in the recoil of the $B_{reco}$ candidates. 
The number of observed $B_{reco}$ events which contain a charged
lepton  with $p^*>$ 1\gevc is denoted as
$N_{sl}^{meas}$. It can be related to the true number of semileptonic
decays, $N_{sl}^{true}$ and the remaining peaking
background $BG_{sl}$, estimated with Monte Carlo simulation, by
$N_{sl}^{true} = (N_{sl}^{meas} - BG_{sl})/\epsilon_l^{sl} \epsilon_t^{sl}=
N_{sl}/\epsilon_l^{sl} \epsilon_t^{sl}$ .
Here $\epsilon_l^{sl}$ refers to the efficiency for selecting a lepton
from a semileptonic B decay with a momentum above $p_{cut}$ in an
event with a reconstructed $B$ with efficiency $\epsilon_t^{sl}$. Figure \ref{fig:mxhadfit} shows the 
result of the \mes\ fit used to determine $N_{sl}^{meas}$. 

If we denote as $N_u^{meas}$ the number of  events fitted in the
sample  after all requirements, and  with $BG_u$ the peaking background
coming from semileptonic decays other than the signal, the true number of signal events $N_u^{true}$
is related to them by
\begin{equation}
N_u= N_u^{meas} - BG_u = \epsilon_{sel}^u  \epsilon_l^u \epsilon_t^u N_u^{true} ,
\end{equation}
\noindent
where the signal efficiency $\epsilon_{sel}^u $ accounts for all selection
criteria applied on the sample after the requirement of 
a high momentum lepton.  

\begin{figure}
 \begin{centering}
 \epsfig{figure=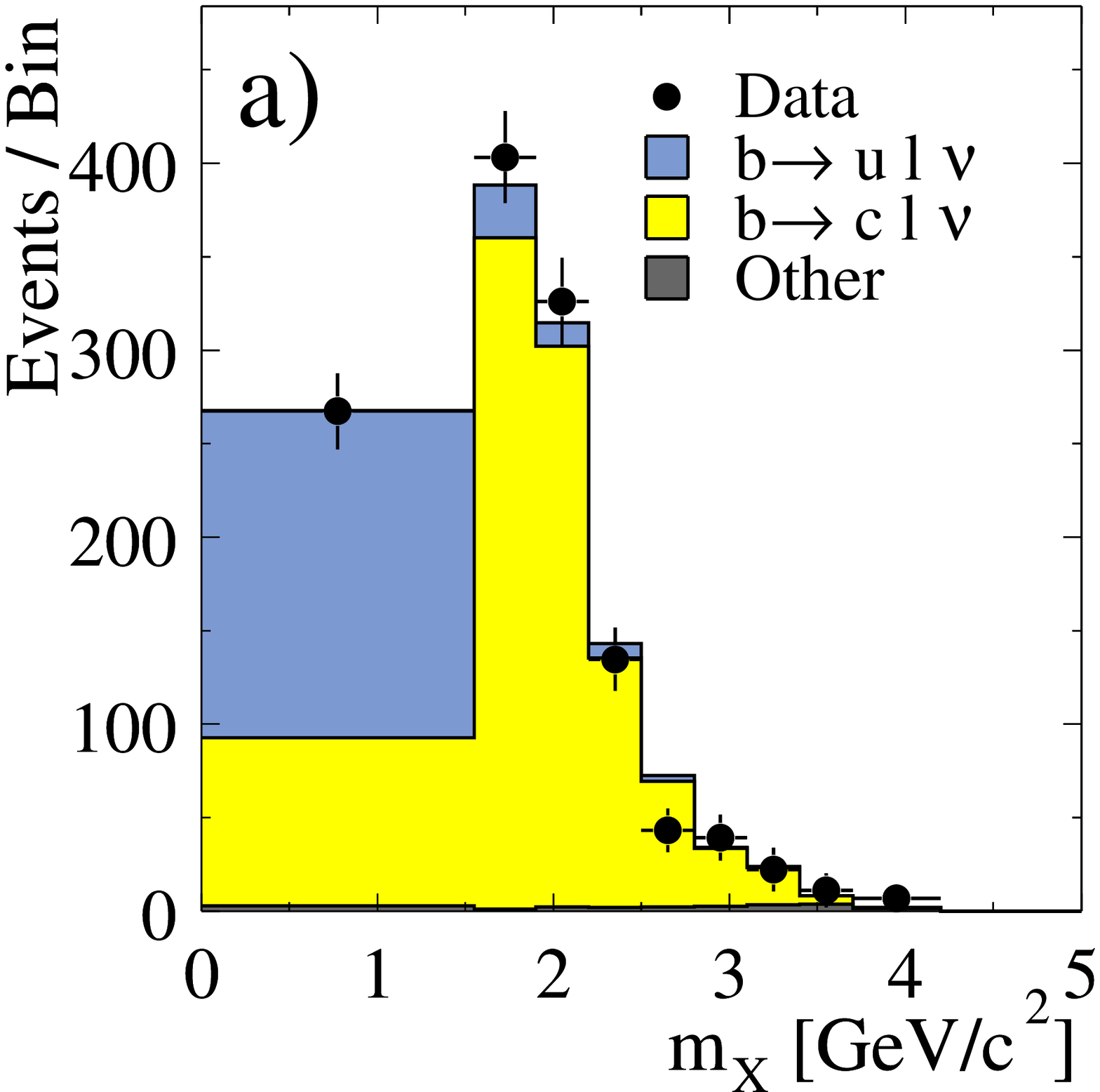 ,height=7cm}
 \epsfig{figure=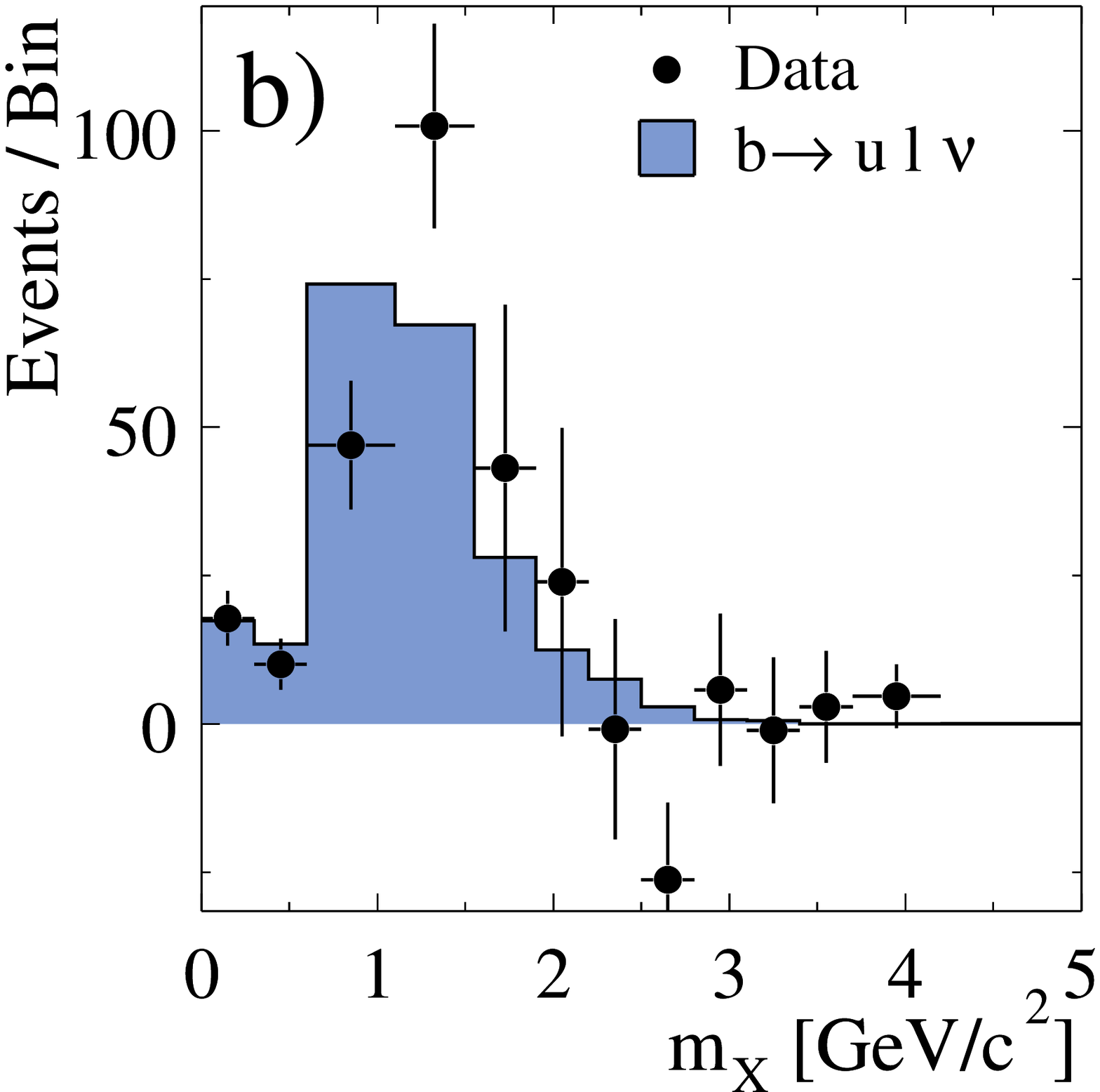 ,height=7cm}
\caption{ $\chi^2$ fit to the \mX\ distribution: a)
     data (points) and fit components, and b) 
     data and signal MC after subtraction of the $b\to c\ell\nu$ and the ``other'' backgrounds.  
\label{mxfit}}
\end{centering}
\end{figure}

To measure $BG_u$ in the inclusive studies, the peaking background ($BG_u$) is estimated by performing a $\chi^2$ fit on 
the \mx\ or \mx-\Q\ distributions, resulting from \mes\ fits in individual \mx\ or \mx-\Q\ bins, 
with the shape of the background estimated from Monte Carlo simulation,
and its normalization free to vary. An 
example of such a $\chi^2$ fit for the \mx\ analysis is shown in Fig.~\ref{mxfit}. 
For the exclusive analysis the background normalization is 
taken from Monte Carlo scaled to the data luminosity. 

The ratio between the branching fractions for the signal and $\Bxlnu$ decays is 
 \begin{equation}
 R_{u/sl} = 
 \frac{\BR({\rm { signal)}}}{\BR(\Bxlnu)}=
 \frac{N_u^{true}}{N_{sl}^{true}} = 
 \frac{(N_u^{meas}- BG_u)/(\epsilon_{sel}^u)}{(N_{sl}^{meas}-BG_{sl})} 
 \times \frac{\epsilon_l^{sl} \epsilon_t^{sl} } {\epsilon_l^u \epsilon_t^u }.
 \label{eq:ratioBR}
 \end{equation}

\noindent
The efficiency ratio is expected to be close to, but not equal to
unity.  Due to the difference in multiplicity and the different
lepton momentum spectra, we expect the tag efficiency $\epsilon_t$
and lepton efficiency $\epsilon_l$ to be slightly different for the
two classes of events.  
The signal branching fraction is then obtained from $R_{u/sl}$ using  
the \Bxclnu\ branching ratio measured in \babar, 
\begin{equation}
\label{eq:semilep}
\BR(\Bxclnu)= (10.61 \pm 0.16 (\rm exp.) \pm 0.06 (\rm theo.))\% ~\cite{Aubert:2004aw}.
\end{equation}
which, given the previous \babar\ measurement for the 
charmless semileptonic branching ratio \cite{Aubert:2003zw}, corresponds to a total 
semileptonic branching fraction of $\BR(\Bxlnu)=(10.83\pm 0.19)\%$.

\subsection{Systematic Uncertainties}
\label{sec:sys}
Most of the systematic uncertainties are common to all analyses.
Uncertainties related to the reconstruction of charged tracks
are  determined  by  removing   randomly  a  fraction  of  tracks
corresponding  to  the uncertainty  in  the  track finding  efficiency
($1.3\%$ for tracks  with transverse momentum
$p_\perp > 0.2\gevc$ and  $2.5\%$ for tracks
with  $p_\perp  <  0.2\gevc$).   The  systematic  error  due  to  the
reconstruction of neutral  particles in the EMC is studied by varying the 
resolution and efficiency to match
those found in control samples in data.

We  estimate the systematic  error due  
to particle  identification by
varying the  electron and kaon identification efficiencies  by $\pm 2\%$
and   the   muon   identification   efficiency   by   $\pm3\%$.    The
misidentification probabilities are varied  by $15\%$ for all particles. 

The uncertainty of the $B_{reco}$ combinatorial background subtraction
is estimated by changing the signal shape function to a Gaussian function instead 
of the empirical function of Ref.~\cite{cry}. 
We evaluated the effect of cross-feed between \Bz\ and \Bp\ decays 
by repeating the analysis with only the \Bz\Bzb or the \Bp\Bm Monte Carlo. 

The impact of the charm semileptonic branching fractions has been estimated by varying each of
the exclusive  branching fractions within  one standard deviation  of the
current world average~\cite{Hagiwara:2002fs}.  Similarly, the branching fractions
of charm mesons have been varied, both for exclusive decays and
for inclusive kaon production. To study the cross-feed among the 
charmless modes we also varied the number of 
charmless exclusive semileptonic decays by $30\%$ for $\Bbar \ra \pi \ell \bar{\nu}$
and $\Bbar \ra \rho \ell \bar{\nu}$, by $40\%$ for $\Bbar \ra \omega \ell \bar{\nu}$ and by 
$100\%$ for the remaining exclusive charmless semileptonic B decays.

In presenting the results the  systematic effects are divided into the following
categories: uncertainties related to detector and reconstruction simulation
(tracking reconstruction, neutral reconstruction, lepton identification,
charged Kaon identification), $\sigma_{det}$; uncertainties related to the theoretical model 
of the non-resonant \btoulnu\ decays, $\sigma_{theo}$;
uncertainties related to other aspects of the \btoulnu\  simulation ( 
resonant signal decay branching fractions, 
$s\bs$ production, hadronization, etc.), $\sigma_{ul\nu}$; 
uncertainties due to the limited available Monte Carlo  statistics ($\sigma_{MCstat}$);
uncertainties related to background simulation and subtraction ($\Bbar \ra
D^{(*,**)} \ell \bar{\nu}$ and $D$ branching fractions, fit to the \mx\ distribution), $\sigma_{bkg}$; 
uncertainties related to the subtraction of the combinatorial background and cross-feed,  
$\sigma_{breco}$. 

\section{One-dimensional \mx\ analysis}
\label{sec:mx}
We  already published a measurement of \Vub\ on
 \Bxulnu\ events with $\mx<1.55\gevcc$~\cite{Aubert:2003zw} and 
subtracting backgrounds with a one-dimensional fit to the \mx\ distribution
(see Fig.~\ref{mxfit}). The values for the shape function parameters used
were $\lbarsf = 0.48 \pm 0.12 \gev$ and $\lonesf = - 0.30 \pm 0.11 \gev^2$
and a correlation of -80\%~\cite{Cronin-Hennessy:2001fk} (see Fig.~\ref{fig:ellipses}).
Since this publication, there have been developments in the theoretical
interpretation. Here we report the updated measurement of \Vub\ obtained
by varying the shape function parameters as stated in Sec.~\ref{sec:theosys}, the new central value being
$\lbarsf=
0.545\gevcc$ and $\lonesf=-0.342$.
The experimental systematic errors considered are the same and correspond to the sum in quadrature of $\sigma_{det}$, $\sigma_{ul\nu}$,
$\sigma_{bkg}$, $\sigma_{MCstat}$, and $\sigma_{breco}$. 
The measured total inclusive branching ratio is
\begin{equation}
\rusl=0.0234\pm 0.0027(\rm stat.)\pm 0.0026(\rm sys.)^{+0.0064}_{-0.0038} (\rm theo.)
\end{equation}
which translates into a total inclusive branching fraction
\begin{equation}
\BR(\Bxulnu)=(2.53\pm 0.29(\rm stat.)\pm 0.28(\rm sys.)^{+0.69}_{-0.41} (\rm theo.))\times 10^{-3}.
\end{equation}
From the relationship between the inclusive branching fraction and \Vub (obtained from Eq. 2 in Ref.~\cite{Gibbons:2004dg} 
having updated the OPE parameters to the measurement of \babar~\cite{Aubert:2004aw})
\begin{equation}
|V_{ub}| = 0.00424 \left( \frac{\BR(\Bxulnu)}{0.002}
\frac{1.604 ps}{\tau_B} \right)^{1/2} 
\times
(1.0 \pm 0.048 (\rm OPE + m_b)).
\label{eq:vubextr}
\end{equation}
Here the average \Bz\ and \Bpm\ lifetime is $\tau_B=1.604\; \rm
ps$~\cite{PDG04}. The ``$OPE+m_b$''
 uncertainty is due to perturbative and non perturbative corrections , 
and to the uncertainty on the $b$-quark mass ($\rm m_b$). We then obtain 
\begin{equation}
 \Vub=(4.77\pm 0.28 (\rm stat.)\pm 0.28 (\rm sys.)^{+0.65}_{-0.39}(\rm theo.)\pm 
0.23({\rm OPE+m_b}))10^{-3},
\end{equation}
consistent within the shape function  errors with the 
published $\Vub =(4.62\pm 0.28 (\rm stat.)\pm 0.28 (\rm sys.) \pm 0.40 
(\rm theo.)\pm 0.26(\rm OPE+m_b))\times 10^{-3}$.

%
\section{Unfolding the \mx\ Spectrum}
%
\label{sec:unfold}
The observed \mx\ spectrum is one of the main results of the \mx\
analysis (see Fig.~\ref{mxfit}). 
In order to convert this into a universal observable, we need to unfold detector and selection effects.
 To increase sensitivity to the true distribution,
the spectrum  is re-binned in 310  \mevcc bins in order to match  the resolution.
From the measured \mx\ spectrum, represented by a vector, \xmeas, we extract
the unfolded spectrum, \xtru. 
The relationship between the two spectra is
$\widehat{A}$ \xtru\ = \xmeas\ \cite{Hocker:1996kb}, where the detector
response matrix $\widehat{A}$
describes the effects of limited efficiency and finite resolution of the
measurement. $\widehat{A}$ is estimated on
signal MC with the hybrid model described in Sec.~\ref{sec:theosys} and with
\lbarsf\ and \lonesf\ set to te CLEO best fit values. To reduce systematic uncertainties the simulation has been refined: 
 to account for the unknown details of  the fragmentation of quarks into hadrons, we divide the sample 
into several categories according to the event multiplicity of charged and neutral particles and adjust the fraction
of events in each category in MC to match the one observed in data.  

 In practice, $\widehat{A}$ is a non-invertible
matrix. Additionally, the uncertainties on the measured spectrum have to be
taken into account adequately. 
We use an unfolding method based on a 
singular value decomposition of the detector response matrix and a suitable
regularization procedure of the unfolded result as described 
in Ref.~\cite{Hocker:1996kb}.

The systematic uncertainties considered are those described in
Sec.~\ref{sec:strategy}.  The systematic effects
on the measured spectrum \xmeas\ are evaluated, each giving rise to a
covariance matrix, $\widehat{C}_k^\mathrm{meas}$, on the measured spectrum.
To propagate the covariance matrices for the measured 
distribution of \mx ($\widehat{C}_k^\mathrm{meas}$) to covariance matrices on
the unfolded spectrum $\widehat{C}_k^\mathrm{unf}$ ,
two different approaches have been taken depending on whether the
response  matrix is affected by the systematic under study.
In the case of a constant response matrix, e.g. for background modelling, 
the observed spectrum is smeared a large number of times
according to $\widehat{C}_k^\mathrm{meas}$, the spectra are unfolded and
the variations of the result in each bin give rise to
$\widehat{C}_k^\mathrm{unf}$.  
When the systematics  affect the response matrix, e.g. for tracking and
neutral efficiencies, the whole measurement is redone with the
appropriate changes in reconstruction and the spread serves for the
determination of $\widehat{C}_k^\mathrm{unf}$.
The covariance matrix $\widehat{C}^\mathrm{unf}$ is computed as a sum of the
individual covariance  matrices arising from the different systematic errors,
$\widehat{C}^\mathrm{unf} = \sum_k \widehat{C}_k^\mathrm{unf}$.

The resulting unfolded spectrum needs to be corrected for a small possible 
bias arising
from the regularization procedure.
To estimate this bias we use 
the unfolded spectrum \xtau, apply the detector response matrix 
$\widehat{A}$ to it and fluctuate the resulting spectrum using the covariance
matrix on the measured spectrum. These spectra are then unfolded and the
bin-by-bin mean deviation from the original \xtau\ is taken as bias.

%
\subsection{Unfolding results}
%

\begin{figure}
\includegraphics[width=0.48\textwidth]{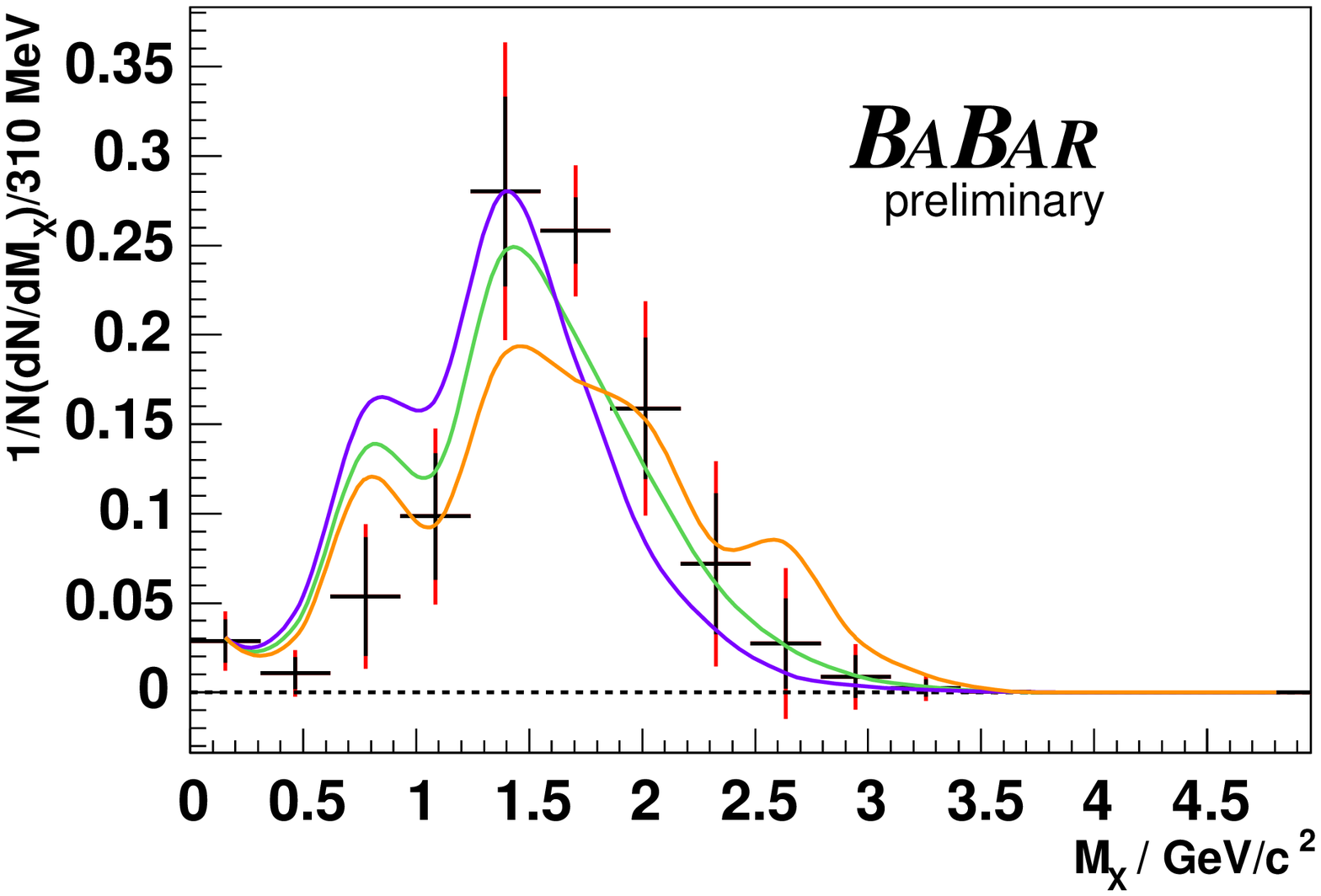} \hfill
\includegraphics[width=0.48\textwidth]{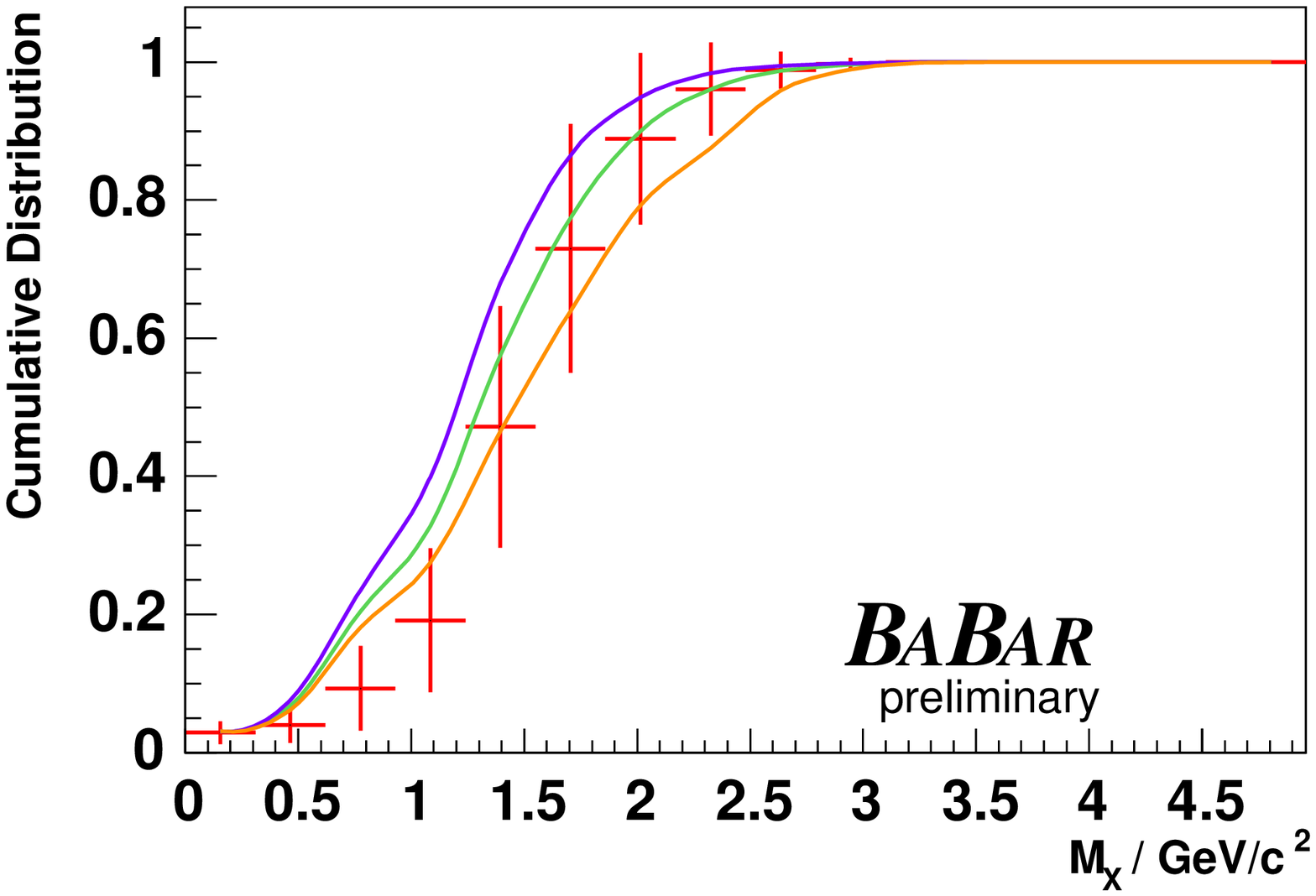}
\caption[Unfolded \mx\ spectrum corrected for bias]{The unfolded spectrum normalized to unit area
(left) and its cumulative distribution (right) as a function of \mx. 
The spectrum and the cumulative distribution from MC with the CLEO best fit $\lbarsf=
0.545\gevcc$ and $\lonesf=-0.342$ is shown in green (middle). 
The orange (highest) and violet (lowest) spectra and cumulative distributions correspond to the two extreme points of the CLEO ellipse 
(see Fig.~\ref{fig:ellipses}),
$\lbarsf=0.800$ and $0.435\gevcc$ and to $\lonesf=-1.22$ and $-0.16$,
respectively. In the left plot 
black errors are only statistical, while the red ones include always systematics. }
\label{unf:result}
\end{figure}

Figure~\ref{unf:result} shows the unfolded spectrum \xtau\ normalized to unit
area and its cumulative. The errors on
the spectrum are given by the square-roots of the diagonal elements of  
the covariance matrix \Ctau. Detailed
listings of \xtau\  and \Ctau\ are given in Table~\ref{tab:spectrum} and~\ref{textcovmat} in  Appendix~\ref{app_tables}.
The uncertainties on the cumulative distribution
take the bin-by-bin correlations of the unfolded spectrum into account.
In  Fig.~\ref{unf:result} we also show a comparison of the data with Monte Carlo samples simulated
with SF parameters corresponding to the CLEO best fit and extremes of the $\Delta\chi^2=1$ contour. The sensitivity of these
spectra to the SF parameters is comparable with the CLEO $b\to s\gamma$ constraint.

Finally, 
we compute the first moment $\mathcal{M}_{1}=\int_0^{M_{X,0}} M_X f(M_X) \d
 M_X/(\int_0^{M_{X,0}} f(M_X) \d M_X)$ 
 and second central moment $\mathcal{M}_{2}^{'}=\int_0^{M_{X,0}} (M_X-\mathcal{M}_{1})^2
 f(M_X) \d M_X/(\int_0^{M_{X,0}} f(M_X) \d M_X)$
from the unfolded spectrum $f(M_X)$, for different upper bounds $M_{X,0}$ on
the \mx\ spectrum. 
Table~\ref{moments} shows the measured moments for 
two different  values of  $M_{X,0}$:  $5\unit{GeV}/c^2$, corresponding to no cut at all, and 
$1.86\unit{GeV}/c^2$. The quoted uncertainties  take the bin-by-bin
correlations of the unfolded spectrum into account. With a higher precision
of the measurement it will be interesting to use 
moments of the hadronic mass spectrum to constrain the shape function
parameters.

\begin{table}[h]
\centering
\caption{First and second central moments of the unfolded \mx\ spectrum. The
correlation between the first and second central moment with the same
$M_{X,0}$  is 
also reported. Finally the breakdown of the error into statistical
uncertainty($\sigma_\upr{stat}\oplus  \sigma_\upr{MCstat}$), uncertainties related to detector effects
($\sigma_\upr{det}$), uncertainties related to signal
modeling ($\sigma_\upr{ul\nu}\oplus \sigma_\upr{theo}$) and to background modeling and subtraction
($\sigma_\upr{bkg} \oplus \sigma_\upr{breco}$) is given in \gevcc and
\gevccsq, respectively. The values of $\mathcal{M}_{1}$ are in \gevcc\ while the values of $\mathcal{M}_{2}^{'}$  are 
in \gevccsq. } 
\renewcommand{\arraystretch}{1.15}
\begin{tabular}{l c c c c c c c c}
\hline \hline 
 &  $M_{X,0}$ (\gevcc) & $\mathcal{M}$ & $\sigma(\mathcal{M})$ & Correlation
 &$\sigma_\upr{stat}\oplus  \sigma_\upr{MCstat}$  &$\sigma_\upr{det}$ & $\sigma_\upr{ul\nu}\oplus \sigma_\upr{theo}$ &
$\sigma_\upr{bkg} \oplus \sigma_\upr{breco}$\\ 
\hline












$\mathcal{M}_{1}$  &$1.86$&$1.355$&$0.084$ & 
&$0.064$&$0.022$&$0.017$&$0.023$ \\  

$\mathcal{M}_{2}^{'}$  &$1.86$&$0.147$&$0.034$&$-0.819$
&$0.027$&$0.010$&$0.005$&$0.011$ \\

$\mathcal{M}_{1}$  &5&$1.584$&$0.233$ & 
&$0.166$&$0.073$&$0.060$&$0.101$ \\

$\mathcal{M}_{2}^{'}$   &5&$0.270$&$0.099$&$0.796$  
&$0.055$&$0.037$&$0.023$&$0.068$ \\

\hline \hline
\end{tabular}
\label{moments}
\end{table}

\section{Two-dimensional \mx-\Q\ analysis}
\label{sec:mxq2}
The \mx\ analysis is systematically limited by the dependence on the 
shape function. This can be overcome by selecting a phase space region where the 
shape function effects are small, namely the region at large
\Q\ values~\cite{Bauer:2001yb}. 
Therefore we find a trade-off between the statistical 
and theoretical uncertainties by loosening the \mx\ cut and applying a
\Q\ one.
 Moreover, since most of the theoretical uncertainties are due to the 
extrapolation from a selected kinematic region to the full phase space,  
measurements of partial branching fractions in different regions of phase space and their extrapolation to the 
full phase space can serve as tests of the theoretical calculations and models.

\begin{figure}
\centerline{\epsfig{figure=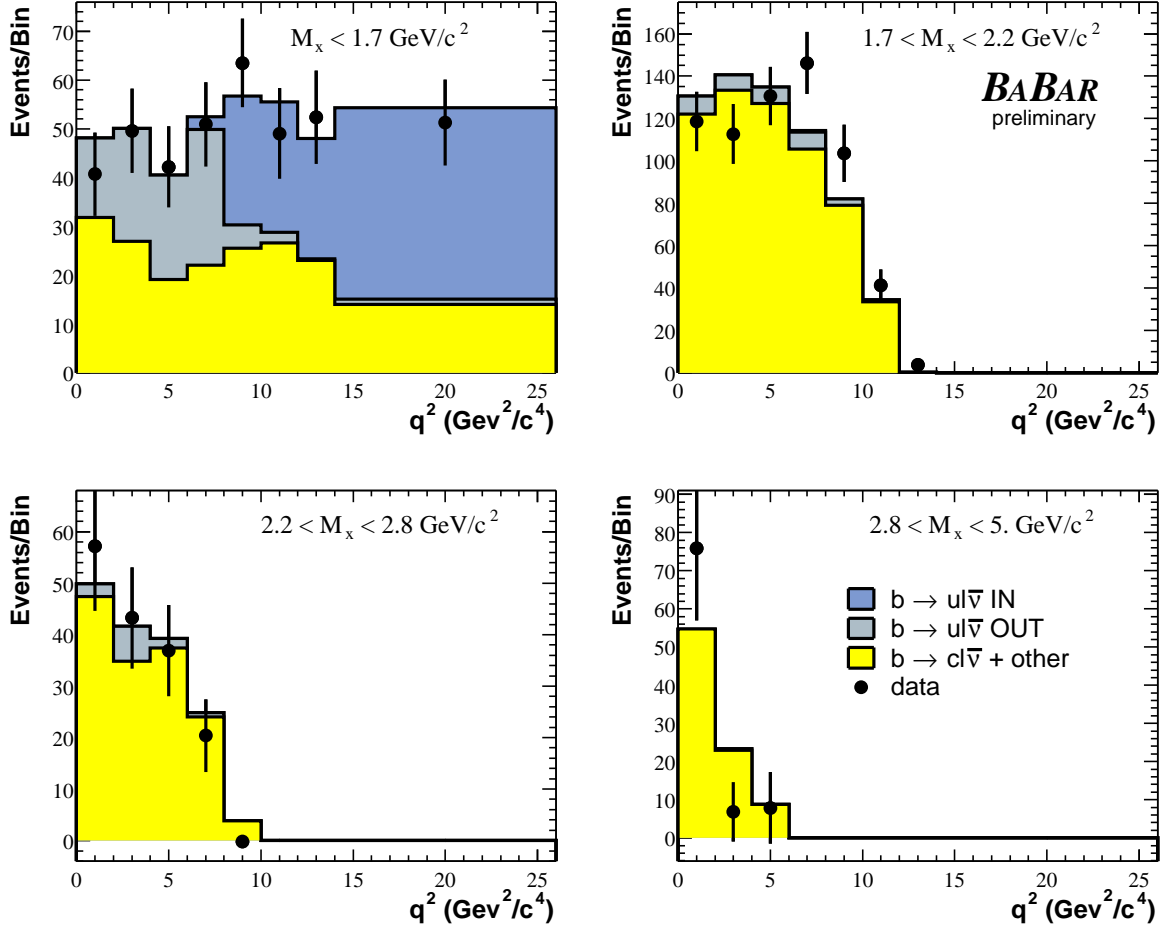,width=16.cm}}
\caption{Distributions of \Q\ in bins of \mx. Points are data, the blue, light gray and yellow histograms  
represent  the fitted contributions from \btoulnu\ events with
true  \mx$<1.7\gevcc$, \Q$>8\gevccsq$ ,  \btoulnu\
events not satisfying these requirements,  and background events, respectively.
}
\label{fig:mxq2fit}
\end{figure}

In order to extract the partial charmless semileptonic branching ratio in a given region of the 
\mx-\Q\ plane $\Delta \BR(\Bxulnu)$,  we define as signal
the events with true values of the kinematic variables in the chosen
region, treating as background those that migrate from outside this region
because of the resolution.
 This means that in applying Eq.~\ref{eq:ratioBR} we include
 the \btoulnu\ events outside the signal region in $BG_u$ and the quoted efficiencies refer only to events 
generated in the chosen (\mx-\Q) region. These efficiencies are computed on Monte Carlo, and therefore are based on 
the DFN model. However, the associated theoretical uncertainty on the final result is small compared to the 
extrapolation error to the full phase space. We divide the events into two-dimensional bins of \mx\ and \Q, 
we fit the \mes\ distribution to extract the yield in each bin, and 
we perform a two-dimensional binned fit of the entire \mx-\Q\ distribution in order to extract the signal and background
 components. The result of the fit 
is shown in Fig.~\ref{fig:mxq2fit}.

\begin{figure}
\centerline{\epsfig{figure=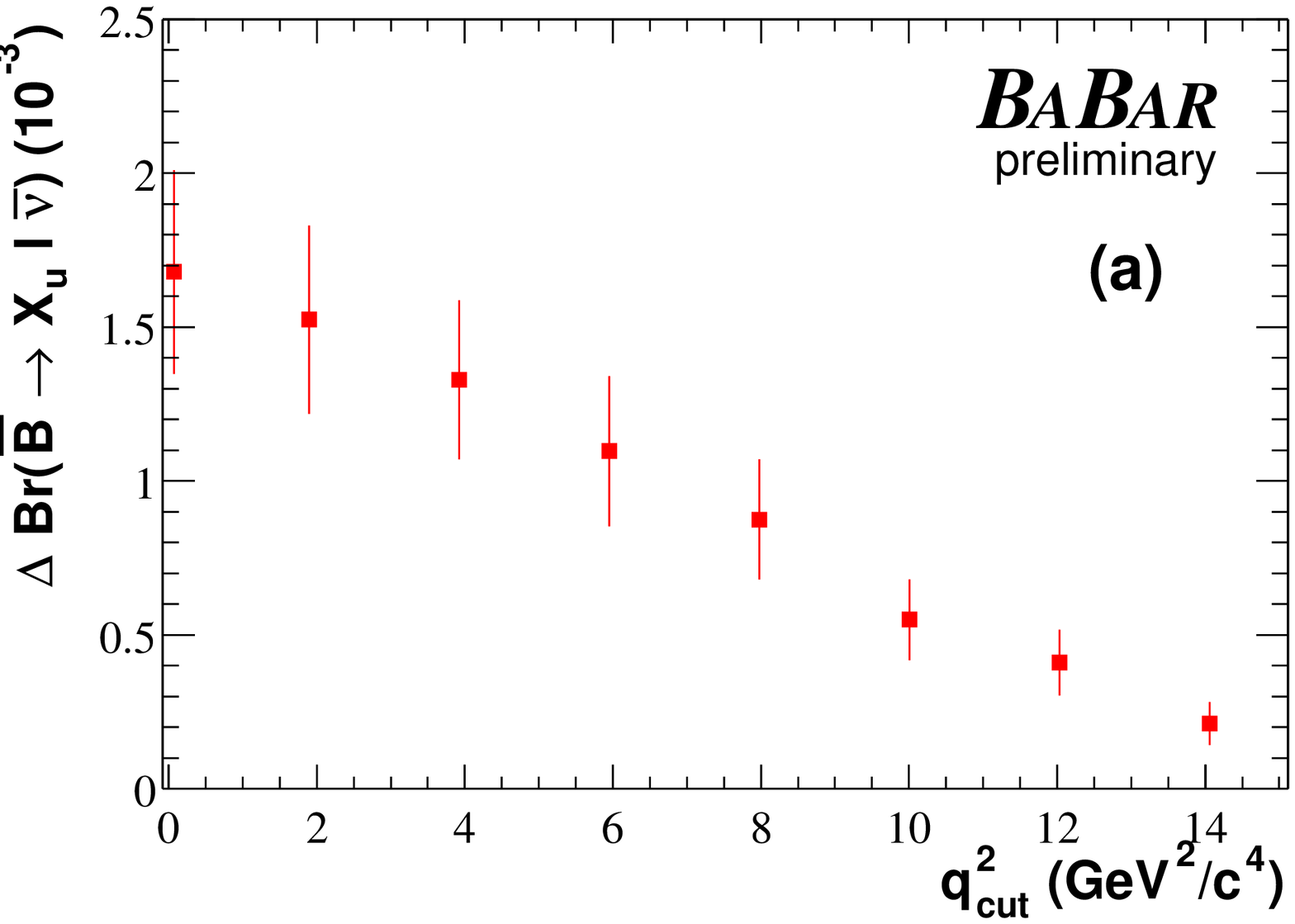,height=7cm}}
\centerline{\epsfig{figure=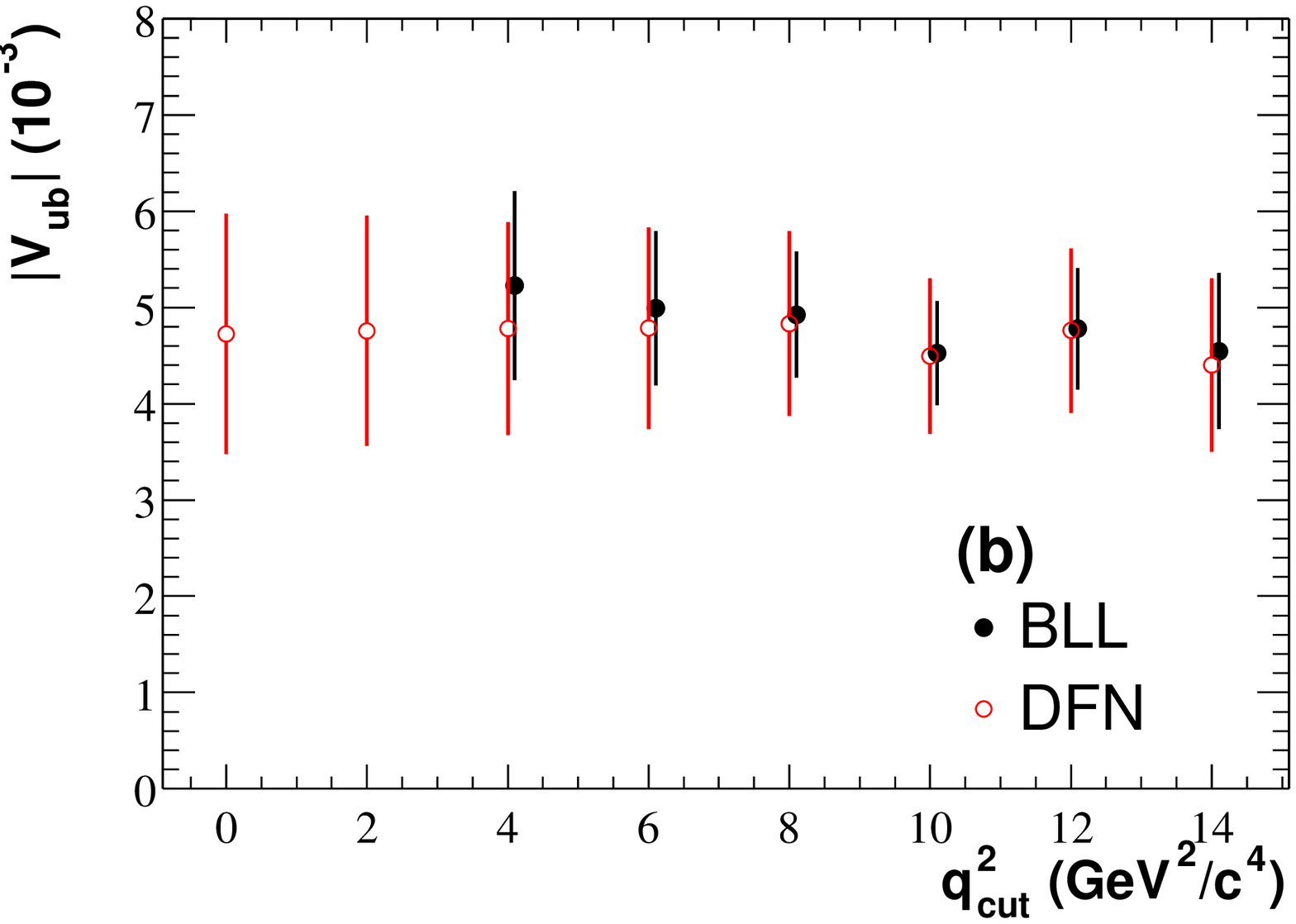,height=7cm}}
\begin{center}
\caption{
(a) Measured partial branching ratio for $\mx<1.7$ \gevcc and $\Q>\Q_{cut}$, as a function of $\Q_{cut}$. 
The error bar is the sum in quadrature 
of statistical, systematical and theoretical uncertainties. 
(b) Measured value of \Vub for $\mx<1.7\gevcc$ as a function of the \Q\ cut
applied when using acceptances from DFN (open points) and BLL
(solid points). The error bars include the statistical, systematic and theoretical 
uncertainties, added in quadrature.}
\label{fig:mxq2dbr}
\end{center}
\end{figure}

 Fig.~\ref{fig:mxq2dbr}a and Table~\ref{tab:partial} show, for $\mx<1.7\gevcc$,
the partial branching fraction $\Delta \BR(\Bxulnu)$ as a function of the minimum \Q\ cut.

\begin{table}[!b]
\begin{center}
\caption{Partial branching fraction measurements (in 10$^{-3}$ units) for 
$\mx<1.7$ \gevcc and $\Q>\Q_{cut}$, as a function of $\Q_{cut}$. 
The different sources of uncertainty (as described in \ref{sec:sys}) are also reported.}
\vspace{0.1in}
\begin{tabular}{|l|c|c|c|c|c|c|c|c|} 
\hline
$\Q_{cut}>$ & $\Delta \BR(\Bxulnu)$ & $\sigma_{stat}$ & $\sigma_{det}$ & $\sigma_{breco}$ & $\sigma_{bkg}$ & $\sigma_{theo}$ & $\sigma_{ul\nu}$ & $\sigma_{MCstat}$ \\ \hline
0     &   1.68    &  0.22 &  0.15 &  0.08 &  0.12 & -0.045 +0.035 &  0.137  &  0.08 \\
2     &   1.52    &  0.20 &  0.16 &  0.07 &  0.11 & -0.028 +0.036 &  0.110  &  0.07 \\
4     &   1.33    &  0.18 &  0.11 &  0.06 &  0.10 & -0.040 +0.026 &  0.116  &  0.06 \\
6     &   1.10    &  0.16 &  0.14 &  0.05 &  0.08 & -0.022 +0.018 &  0.083  &  0.05 \\
8     &   0.88    &  0.14 &  0.09 &  0.04 &  0.06 & -0.028 +0.009 &  0.053  &  0.05 \\
10    &   0.55    &  0.11 &  0.03 &  0.02 &  0.04 & -0.006 +0.019 &  0.027  &  0.03 \\
12    &   0.41    &  0.09 &  0.04 &  0.02 &  0.03 & -0.010 +0.000 &  0.033  &  0.03 \\
14    &   0.21    &  0.06 &  0.01 &  0.01 &  0.02 & -0.012 +0.012 &  0.018  &  0.02 \\

\hline
\end{tabular}
\label{tab:partial}
\end{center}
\end{table}

We convert the measured $\Delta \BR(\Bxulnu)$ into \Vub\ by
\begin{equation}
\label{eq:dbrvub}
 |V_{ub}| = \sqrt{\frac{192 \pi^3}{\tau_B G_F^2 m_b^5}\frac{\Delta \BR(\Bxulnu)}{G}} 
\end{equation}
where $\tau_B = 1.61 ps$ and $G$ is a theoretical parameter calculated in the BLL approach~\cite{Bauer:2001yb}. 
The first factor under the square root is 
192$\pi^3/(\tau_B G_F^2 m_b^5)=0.00779$. 
The measured \Vub\ as a function of the \Q\ cut is shown in
Fig.~\ref{fig:mxq2dbr}b for the acceptances computed by BLL and by
DFN. Note that, since the operator product expansion breaks down when going to low \Q, the BLL calculation is only possible for higher values of
\Q.

The error on the acceptance as computed by BLL increases for tighter cuts on 
\Q. For smaller values of \Q, the shape function effects increase. 
In the signal region $\Q>8\gevccsq$, $\mx<1.7\gevcc$
we obtain: 
\begin{equation}
 \Delta \BR(\Bxulnu,\mx<1.7 \gevcc, \Q>8  \gevccsq) = (0.88 \pm 0.14 (\rm stat.) \pm 0.13 (\rm sys.) \pm 0.02 (\rm theo.)) \times 10^{-3}. 
\label{mxq2res}
\end{equation}
To extract \Vub, we take $G$ as computed by BLL and rescale it to the 
$b$-quark mass as measured by \babar\cite{Aubert:2004aw}, 
obtaining $G = 0.282 \pm 0.053$, corresponding to an acceptance $\epsilon_{BLL}=0.325\pm 0.061$. 
Eq.~\ref{eq:dbrvub} yields
\begin{equation}
|V_{ub}|  = (4.92 \pm 0.39 (\rm stat.) \pm 0.36 (\rm sys.) \pm 0.46 (\rm theo.)) \times 10^{-3}.
\end{equation}

In the DFN model the calculated acceptance is $\epsilon =  0.337^{+0.037}_{-0.074}$
and by using Equation~\ref{eq:vubextr} we obtain 
$\Vub= (4.85 \pm 0.39 (\rm stat.) \pm 0.36 (\rm sys.) $ $^{+0.54}_{-0.29} (\rm theo.)) \times 10^{-3}$, in  
agreement with the extraction based on BLL, as well as with the result form the one-dimensional \mx\ fit. 
Figure~\ref{fig:mxq2dbr} shows the measured values for \Vub\ as a function of the \Q\ cut for 
$\mx<1.7$ \gevcc, showing good consistency between the different cuts and theoretical framework. 
Checks were done also with a looser ($\mx<1.86$ \gevcc) 
and a tighter ($\mx<1.5$ \gevcc) 
cut on \mx, and they give consistent results. 

\section{Results based on Belle's estimate of the SF parameters} 
\label{sec:belle}
We report here the results obtained with the SF parameters as estimated from the  $b\ra s \gamma$ photon 
energy spectrum measured 
by Belle (see Sec.~\ref{sec:theosys}) and we reinterpret our results by using them in the \mx\ and \mx-\Q\ analyses. 
The  acceptance obtained for the DFN model is lower, and therefore
 the charmless semileptonic branching fraction and  \Vub\ are higher. 
The theoretical systematics due to the shape function parameters is reduced, due to the 
significantly better precision.

For the \mx\ analysis we get 

\begin{equation}
\rusl=(2.81\pm 0.32(\rm stat.)\pm 0.31(\rm sys.)^{+0.23}_{-0.21} (\rm theo.))\times 10^{-2}
\end{equation}

which translates into 

\begin{equation}
 \Vub=(5.22\pm 0.30(\rm stat.)\pm 0.31 (\rm sys.)^{+0.22}_{-0.20}(\rm SF)\pm 
0.25(\rm pert+1/mb^3))10^{-3}.
\end{equation}
\begin{figure}
\includegraphics[width=0.48\textwidth]{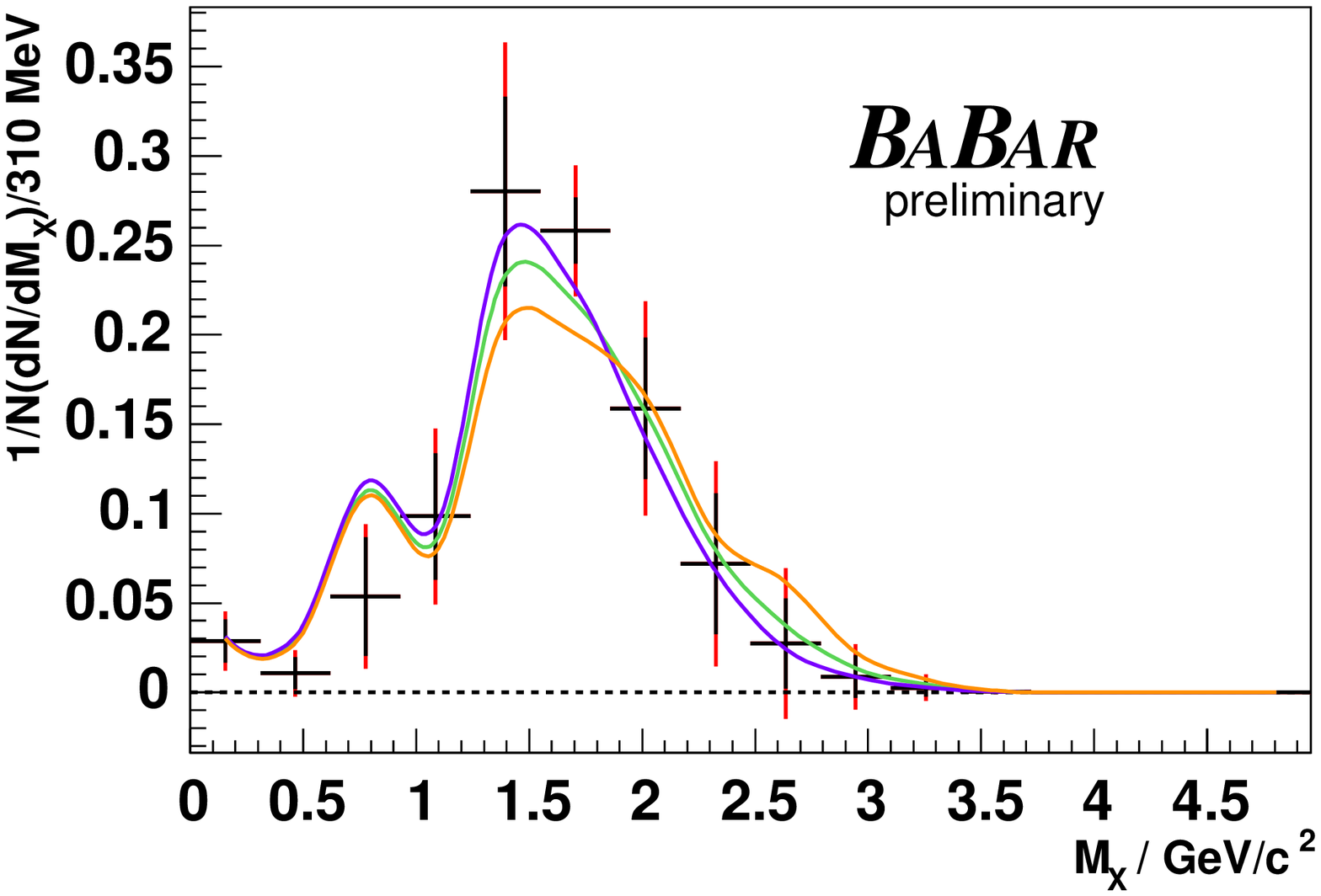} \hfill
\includegraphics[width=0.48\textwidth]{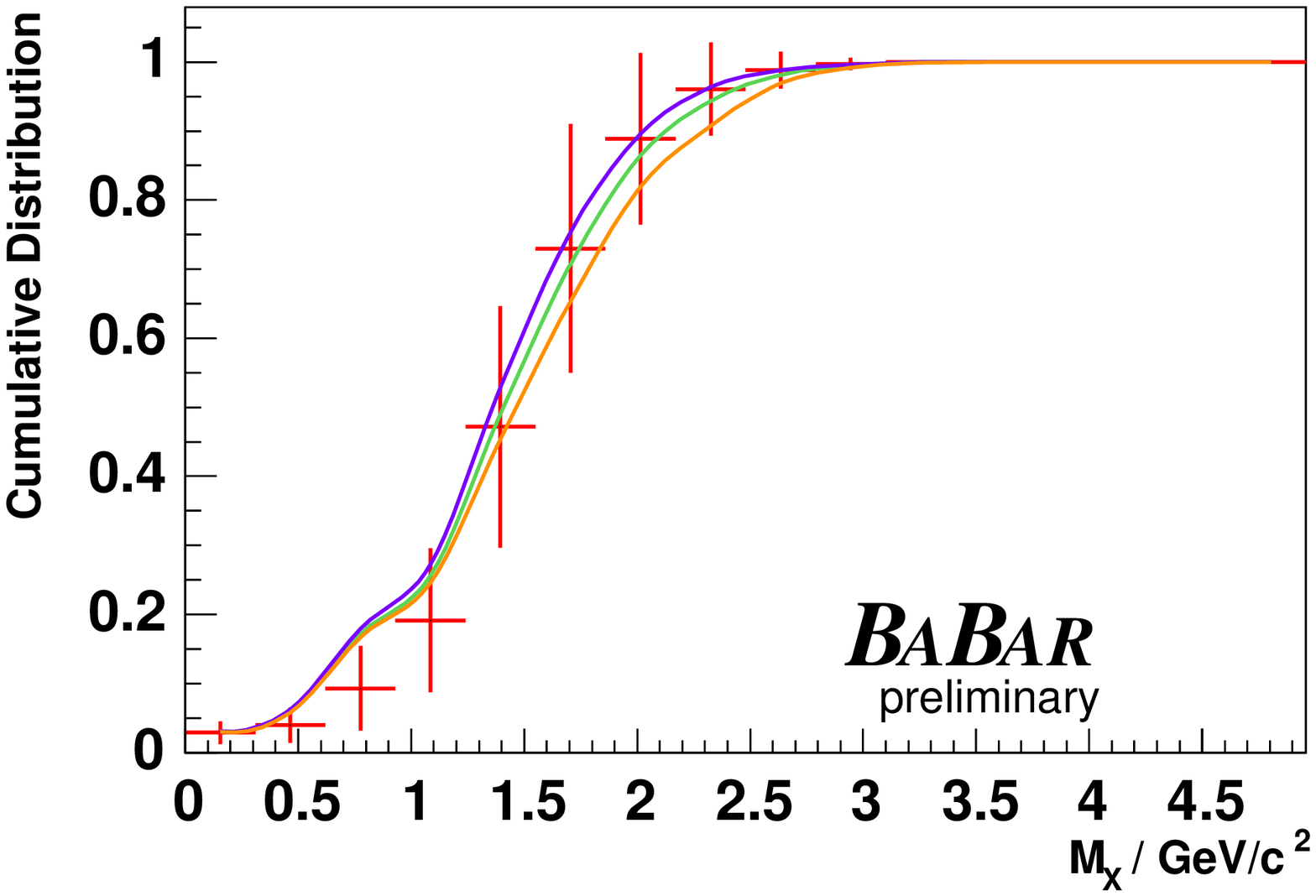}
\caption[Unfolded \mx\ spectrum corrected for bias]{The unfolded spectrum
(left) and its cumulative distribution (right) as a function of \mx. 
The spectrum and the cumulative distribution from MC with the Belle best fit $\lbarsf=
0.66\gevcc$ and $\lonesf = -0.4\gevccsq$ is shown in green. 
The orange and violet spectra and cumulative distributions correspond to the two extreme points in the 
Belle ellipse (see Fig.~\ref{fig:ellipses}), $\lbarsf=
0.600\gevcc$ and  $\lbarsf=
0.748\gevcc$ and to $\lonesf = -0.66(\gevcc)^2$  and  $\lonesf =
-0.28(\gevcc)^2$, respectively. 
In the left plot 
black errors are only statistical, while the red ones always include systematics.}
\label{unf:result2}
\end{figure}

As far as the unfolding is concerned, 
Fig.~\ref{unf:result2} compares the measured spectra with the
 distributions corresponding to the SF parameters measured by Belle.

The  partial branching fraction measurements as a function of the \Q\ cut obtained by the 
\mx-\Q analysis are reported, for
\mx$<1.7$ \gevcc, in Table~\ref{pbfbelle}. 

\begin{table}[!b]
\begin{center}
\caption{BELLE ellipse: Partial branching fraction $\Delta\BR(\Bxulnu)$ measurements (in 10$^{-3}$ units) 
for different \Q\ cuts. \mx\ is required to be less than 1.7\gevcc. 
The different sources of uncertainties are also reported.}
\vspace{0.1in}
\begin{tabular}{|l|c|c|c|c|c|c|c|c|} 
\hline
$\Q_{cut}>$ & $\Delta \BR(\Bxulnu)$ & $\sigma_{stat}$ & $\sigma_{det}$ & $\sigma_{breco}$ & $\sigma_{bkg}$ & $\sigma_{theo}$ & $\sigma_{ul\nu}$ & $\sigma_{MCstat}$ \\ \hline
0  & 1.740 &  0.231 &  0.159 &  0.078 &  0.129 & -0.042 +0.026 &  0.158 &  0.078 \\
2  & 1.584 &  0.205 &  0.165 &  0.071 &  0.117 & -0.037 +0.027 &  0.145 &  0.068 \\
4  & 1.381 &  0.186 &  0.113 &  0.062 &  0.102 & -0.036 +0.026 &  0.140 &  0.061 \\
6  & 1.135 &  0.161 &  0.144 &  0.051 &  0.084 & -0.025 +0.017 &  0.105 &  0.053 \\
8  & 0.896 &  0.143 &  0.091 &  0.040 &  0.066 & -0.017 +0.012 &  0.064 &  0.047 \\
10 & 0.566 &  0.113 &  0.026 &  0.025 &  0.042 & -0.006 +0.013 &  0.041 &  0.036 \\
12 & 0.406 &  0.085 &  0.038 &  0.018 &  0.030 & -0.002 +0.003 &  0.034 &  0.026 \\
14 & 0.207 &  0.059 &  0.014 &  0.009 &  0.015 & -0.007 +0.002 &  0.026 &  0.019 \\
\hline
\end{tabular}
\label{pbfbelle}
\end{center}
\end{table}

The measurement of the partial branching fraction $\BR(\Bxulnu)$ 
in the region limited by 
$\mx < 1.7 \gevcc, ~~~~ \Q>8 (\gevcc)^2$ is 
\begin{eqnarray*}
\Delta\BR(\Bxulnu,\mx<1.7 \gevcc, \Q>8 \gevccsq)=(0.90 \pm 0.14(\rm stat.) \pm 0.14(\rm sys.)^{+0.01}_{-0.02}(\rm theo.))\times 10^{-3}.\\
\end{eqnarray*}

By using $G = 0.282 \pm 0.053$ from BLL, we get 
\begin{eqnarray*}
|V_{ub}| & = & (4.98 \pm 0.40(\rm stat.) \pm 0.39(\rm syst.) \pm 0.47(\rm theo.)) \times 10^{-3}. \\
\end{eqnarray*}
The DFN acceptance computed at \mx$<1.7$ \gevcc\ and \Q$>8$ (\gevcc)$^2$ with the Belle SF parameters is 
$\epsilon =  0.300^{+0.023}_{-0.028}$. This gives in the DFN framework 
$|V_{ub}| = (5.18 \pm 0.41_{stat} \pm 0.40_{syst} ~^{+0.25}_{-0.20~~~theo}) \times 10^{-3}$.

\begin{figure}[!t]
\begin{centering}
\centerline{\epsfig{figure=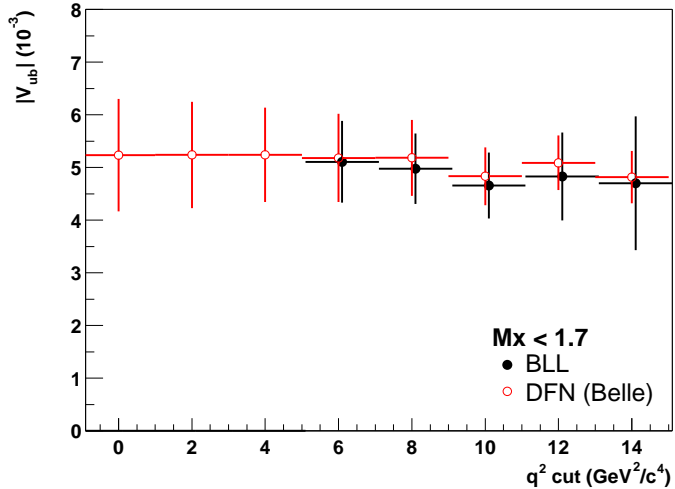,height=7cm}}
\caption{BELLE ellipse: Values for $|$V$_{ub}|$ as a function of the \Q\ cut for \mx$<1.7$ \gevcc\ by taking 
acceptances from Bauer, Ligeti and Luke (points) and De~Fazio-Neubert (squares). The error is the sum 
in quadrature of the statistical, systematic and theoretical uncertainties.}
\label{vubscanbelle}
\end{centering}
\end{figure}

Figure~\ref{vubscanbelle} shows the results for \Vub as a function of the \Q\ cut for \mx$<1.7$ \gevcc, for both DFN and BLL. The 
two models are still consistent within the present accuracies. 
The stability of the result and the agreement between the two methods seems to indicate that OPE is still valid in this \Q\ range.

\section{Measurement of Exclusive Charmless Semileptonic Branching Fractions}
\label{sec:excl}
\subsection{Reconstruction of Exclusive Modes.}
\label{sec:exclreco}
The reconstruction and selection of the events follows closely the one
described in Sec.~\ref{sec:strategy}.
Hadrons are reconstructed in the following modes:
\begin{itemize}
 \item \piz candidates are defined as pairs of photons with an energy in the
laboratory $E_{\gamma}>30\mev$. 
In order to further reject combinatorial background 
we apply a cut on the energy of the more energetic photon in the recoiling $B$
meson rest frame at $E^*_{\gamma}>300\mev$.
 \item \rhoz candidates are reconstructed using pairs of charged tracks
   with opposite charge assuming the pion mass.
 A cut on the momentum of the pions in the recoiling B meson rest frame is applied. The tracks are
ordered depending on their momentum and we apply a cut on the more energetic
track at $p_{\pi}^*>350\mevc$ and on the other one at $p_{\pi}^*>150\mevc$.
 \item \rhop candidates are reconstructed using one charged track
(with opposite charge with respect to the most energetic lepton in the recoil)
and one \piz.
 We  apply a cut on the momentum of the \piz momentum
at $p_{\piz}^*>150\mevc$.
 \item $\omega$ candidates are reconstructed in the mode $\omega \rightarrow \pip\pim\piz$
($\BR(\omega \rightarrow \pip\pim\piz) = 89.1\%$) by using pairs of charged
tracks with opposite charge and one \piz. In 
order to further reject combinatorics we apply a cut rejecting combinations that are at the edge of 
 the Dalitz plot of the three pions.
 \item $\eta$ candidates are reconstructed in three decay modes: $\eta \rightarrow \gamma \gamma$
($BF = 39.4\%$), $\eta \rightarrow \pip \pim \piz$ ($BF = 22.6\%$) and
$\eta \rightarrow \piz \piz \piz$ ($BF = 32.5\%$).
 \item $\etapr$ candidates are reconstructed in two decay modes:  $\etapr \rightarrow \rho \gamma$
($BF = 29.5\%$) and $\etapr \rightarrow \eta \pip \pim$ ($BF = 44.3\%$). We apply cuts on the
mass of the daughters at $0.508\gevcc<M_{\eta}<0.588\gevcc$ and  $0.625\gevcc<M_{\rhoz}<0.925\gevcc$.
 \item \az and \azp candidates are reconstructed in the  $\eta \piz$
   mode.
$\eta$'s are required to have a reconstructed mass within the range $0.518\gevcc<M_{\eta}<0.578\gevcc$.  
We assume a Breit-Wigner width of $71 \pm 7 \mevcc$ \cite{Teige:1996fi} for both \az and \azp .
\end{itemize}

In addition to these criteria we also select mass windows around the
nominal meson masses as detailed in Table~\ref{tab:cuts}

\subsection{Event based cuts}
\label{sec:exclevcuts}
The event variables 
 used to reject background are the same as described in Sec.~\ref{sec:strategy}, 
but the cut values are different and have been optimized mode by mode. Also, the
missing  mass squared (\mmiss) is 
computed from the lepton and just the  charged tracks and photon candidates that 
form one of the specific hadronic final states we are looking for. This means that if
more than one mode is reconstructed in a given event, more than one
value of \mmiss\ is calculated.
For each decay mode the best candidate in an event is chosen on the
basis of the $\chi^2$ built from the reconstructed invariant mass of the
resonance daughter and \mmiss.
In addition 
we have applied discriminating variables:
the total photon energy ($E_{neutral}$), the 
 invariant mass of the lepton and the charged track ($m_{trk}$), the 
 number of reconstructed \piz and charged tracks in the recoil.
A summary of selection criteria per mode is shown in Table~\ref{tab:cuts}. 
\begin{table}[!hp]
\begin{center}
\caption{Summary of event based cuts and efficiency for all modes. The 
         calculation of the efficiency does not include hadronic B reconstruction
         efficiency ($\epsilon_t$).}
\vspace{0.1in}
\begin{tabular}{|l|l|l|l|} 
\hline
Common cuts & \multicolumn{3}{c|}{$p^* > 1.0\gevc$, $N_{lepton} = 1$, $Q_{b(recoil)} Q_{\ell} >0$ ,}\\
            & \multicolumn{3}{c|}{$ Q_{tot}= 0$, $N_{K}=0$,} \\
            & \multicolumn{3}{c|}{no additional charged track in the recoil} \\\cline{1-4}
\hline\hline
  &  \bpi & \bpiz & \brhop \\
\hline
\mmiss(\gevccsq)             &  $|\mmiss(\pip)|<0.3$ & $-0.5<\mmiss(\piz)<0.7$ & $|\mmiss(\rho^+)|<0.4$ \\ 
Mass (\gevcc)               &  - & $0.11<m<0.16$ & $0.55<m<1.0$  \\
Specific cuts              & $ E_{neutral}<0.25\gev $  & - & - \\ 
                          & $ |m_{trk}-3.1|<0.02\gevcc$ &   &   \\ \hline
Final eff.($\epsilon_{sel}  \epsilon_l$)  & $0.302 \pm 0.005$ & $0.263 \pm 0.007$ & $0.154 \pm 0.004$ \\ 
\hline\hline
  &  \brhoz & \bomega & \bet\\
\hline
\mmiss(\gevccsq)             &  $|\mmiss(\rho^0)|<0.4$ & $|\mmiss(\omega)|<0.4$ &  $|\mmiss(\eta)|<0.5$ \\ 
Mass (\gevcc)               &  $0.6<m<1.0$ & $0.74<m<0.82$ &  $0.515<m<0.575$  \\
Specific cuts              &  $N(\piz)<2$           & $N(\piz)<6$  & $\mmiss(\piz)<-1.5\gevccsq$ \\ 
                          &  $\mmiss(\omega)<-0.4\gevccsq$ &   &  \\ \hline
Final eff.($\epsilon_{sel}  \epsilon_l$)  & $0.214 \pm 0.006$ & $0.109 \pm 0.004$ & $0.133 \pm 0.006$ \\ 
\hline\hline
  &  \betp & \baz & \bazp  \\
\hline
\mmiss(\gevccsq)             & $|\mmiss(\eta^\prime)|<0.5$ & $|\mmiss(a_0^0)|<0.5$ & $|\mmiss(a_0^+)|<0.5$ \\ 
Mass (\gevcc)               & $0.92<m<0.99$ & $0.92<m<1.04$ & $0.92<m<1.04$ \\
Specific cuts              & & $\mmiss(\omega)>1.5\gevccsq$ & $\mmiss(\rhop)>1.0\gevccsq$ \\ 
                          &   & $\mmiss(\eta)>0.5\gevccsq$   & \\ \hline
Final eff.($\epsilon_{sel}  \epsilon_l$)  & $0.066 \pm 0.004$ & $0.049 \pm 0.008$ & $0.074 \pm 0.007$ \\ 
\hline
\end{tabular}
\label{tab:cuts}
\end{center}
\end{table}

\subsection{Systematic Uncertainties}
\label{sec:exclsyst}

The systematics are summarized in Table~\ref{tab:systematics3}. The sources common to the other
analyses
are described in Sec.~\ref{sec:strategy}, but
there are also sources of systematic uncertainty which are specific to 
this exclusive analysis.

We evaluate the impact of different form factor calculations changing 
the ISGW2 model (our default) to light cone sum rule calculations by reweighting events. This error is included in $\sigma_{ul\nu}$.

The systematic effect due to the non-resonant structure has been evaluated 
by using a different mixture that allows 
the non-resonant contribution to go down to $2 \pi$ masses (the same model used in the 
inclusive analysis~\cite{Aubert:2003zw}).
The \brhoz channel has a different treatment since the $\pip\pim$ non-resonant contamination
may not be negligible and it cannot be distinguished from the 
signal component. In order to quantify the non-resonant contribution 
we use an approach that is very similar to the one described in
\cite{Athar:2003yg}. From measurements of $e^+e^-$ data and $\tau$ decays and from Bose symmetry 
considerations the non-resonant contribution has to come from the isospin $I=0$ component.  
The isospin relationships therefore predict no non-resonant component in 
 \brhop, and a ratio 2:1 between the non resonant contribution due to 
 $B^-\rightarrow\pi^0\pi^0 \ell \bar{\nu}$ and  $B^-\rightarrow\pi^+\pi^- \ell \bar{\nu}$.
By measuring the  $B^-\rightarrow\pi^0\pi^0 \ell \bar{\nu}$ from our data we obtain
\begin{equation}
 \BR( \Bm\rightarrow\piz\piz\ell\bar{\nu})<0.6 \times 10^{-4} \quad (\mbox{90 \% C.L.})
\end{equation}  
in the mass region $0.6\gevcc<m(\piz\piz)<1.0\gevcc$ .       
The 68\% C.L. upper limit is used to evaluate the systematic uncertainty due to the non-resonant 
component. In the following we will quote $\BR(\brhozv)$, assuming no non-resonant contribution since we measured 
a rate for $B^-\rightarrow\pi^0\pi^0 \ell \bar{\nu}$ which is compatible with zero. 
On the other hand this systematic uncertainty is taken into account in the 
combination of the \brhop, \brhoz and \bomega results where isospin and quark model 
relations are used (see next section).
This component is included in $\sigma_{bkg}$. 

\begin{table}[!hpt]
\caption{Summary of the systematic uncertainties in the measurement of
the $R_{u/sl}$ defined in Eq.~\ref{eq:ratioBR}.}
\begin{center}
\begin{tabular}{|l|c|c|c|c|c|}  
\hline 
&\multicolumn{5}{c|}{Uncertainty on $R_{u/sl}[\times 10^{-4}]$} \\\cline{2-6}
                                   & \bpi & \bpiz & \brhop & \brhozv & \bomega \\ 
\hline 

$R_{u/sl}[\times 10^{-4}]$               &0.86 &0.81   &3.3   &0.92  &1.12\\
$\sigma_{stat}$                    &0.33 &0.25   &1.1   &0.34  &0.49\\
\hline
$\sigma_{det}$                     &0.015 &0.063 &0.250 &0.026  &0.076\\
$\sigma_{breco}$                   &0.090 &0.050 &0.257 &0.071  &0.074\\
$\sigma_{bkg}$                     &0.043 &0.046 &0.475 &0.047  &0.064\\
$\sigma_{ul\nu}$                   &0.029 &0.028 &0.234 &0.059  &0.078\\
$\sigma_{MCstat}$                  &0.048 &0.065 &0.257 &0.092  &0.146  \\
\hline                                                               
Total syst. error                  &0.115 &0.117 &0.652 &0.141  &0.207 \\

\hline         
\hline         
                                   & \bet & \betp & \baz & \bazp &  \\ 
\cline{1-5}
$R_{u/sl}[\times 10^{-4}]$               &0.34   &2.4   &1.8   &0.5   &\\
$\sigma_{stat}$                    &0.36   &1.0   &1.0   &1.3   & \\
\cline{1-5}                                                      
$\sigma_{det}$                     &0.167  &0.199 &0.754 &0.081 & \\
$\sigma_{breco}$                   &0.046  &0.122 &0.106 &0.133 & \\
$\sigma_{bkg}$                     &0.038  &0.106 &0.072 &0.047 & \\
$\sigma_{ul\nu}$                   &0.004  &0.034 &0.180 &0.050 & \\
$\sigma_{MCstat}$                  &0.073  &0.290 &0.293 &0.204 & \\
\cline{1-5}                                                      
Total syst. error                  &0.192  &0.389 &0.839 &0.266 &\\
\hline
\end{tabular}
\end{center}
\label{tab:systematics3}
\end{table}

\subsection{Results of the exclusive measurements.}
\label{sec:exclresults}

\begin{figure}
 \begin{centering}
 \epsfig{file=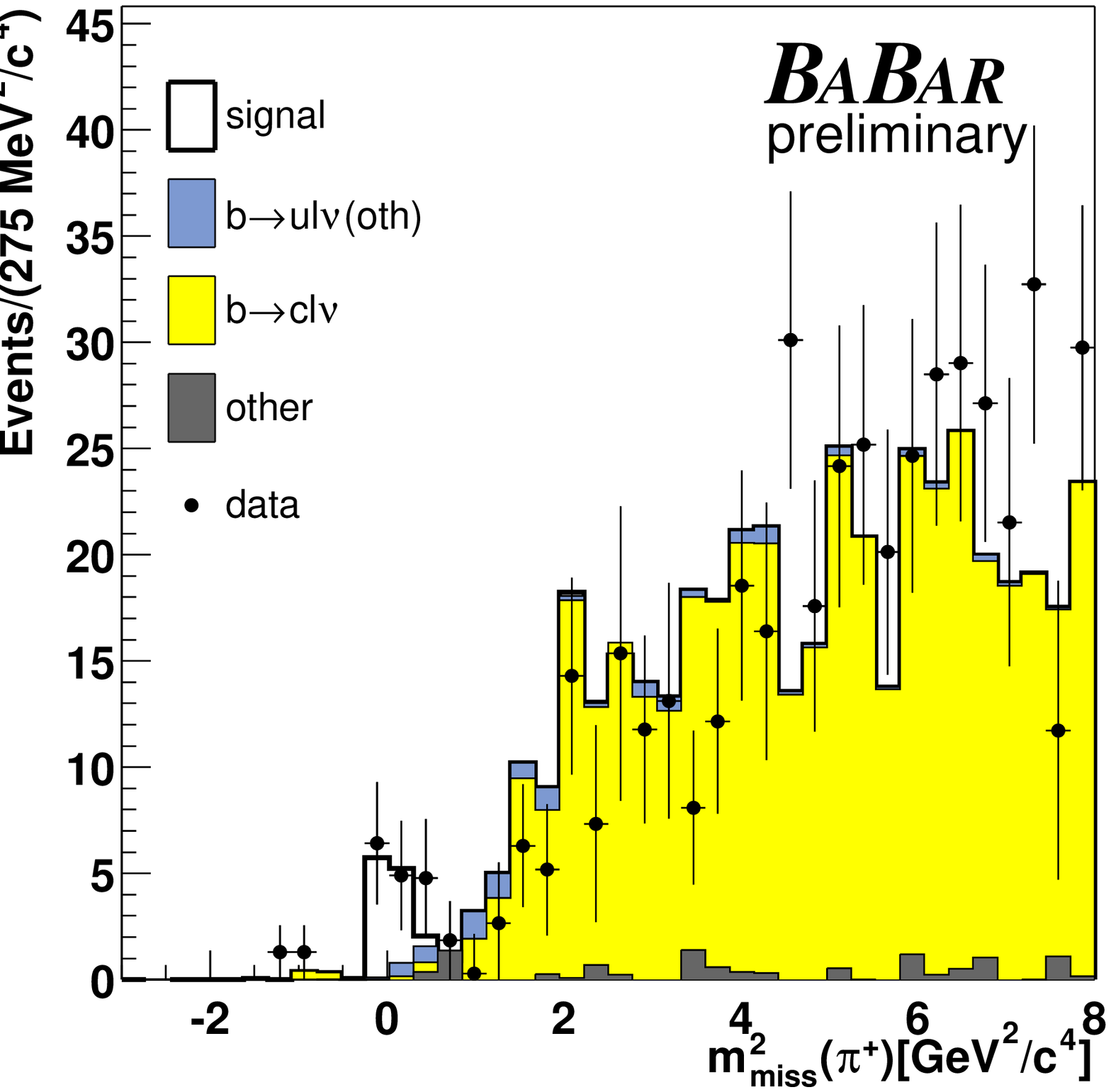,width=8.cm} 
 \caption{ \bpi: Projection of the fit result onto the \mmiss\ variable.
\label{fig:datapifit}}
 \end{centering}
\end{figure}

\begin{figure}
 \begin{centering}
 \epsfig{file=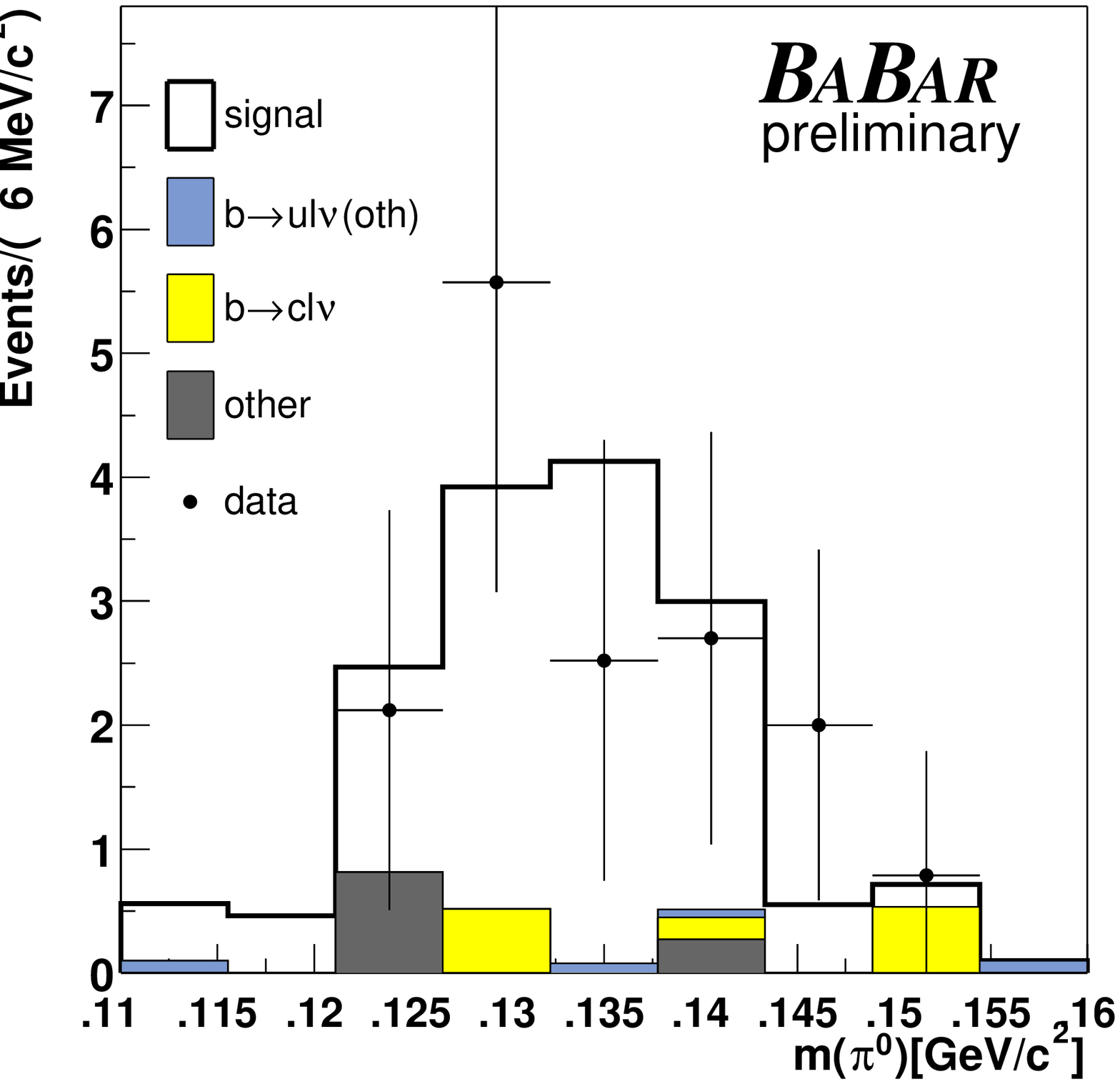,width=8.cm}        
 \epsfig{file=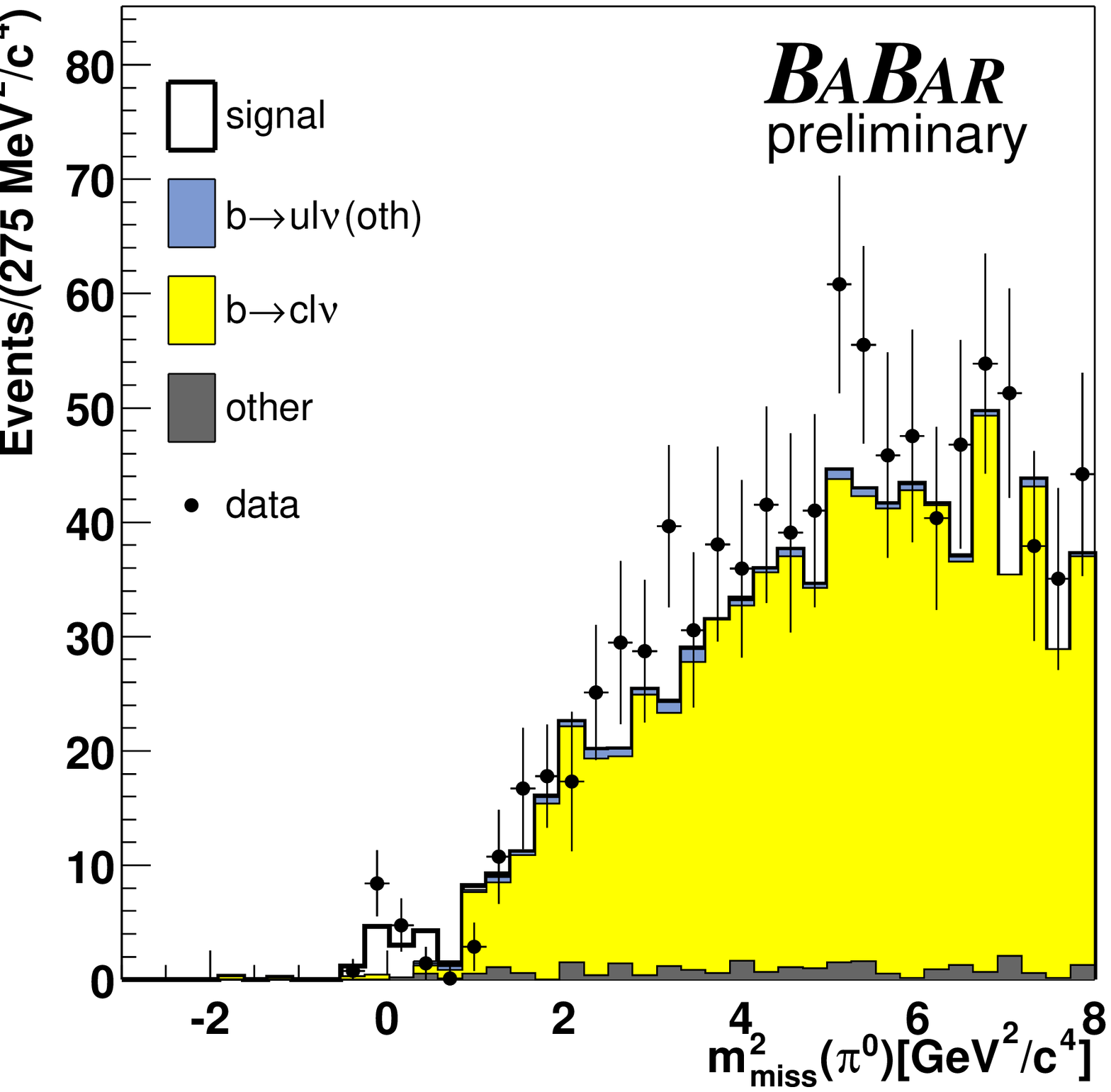,width=8.cm}         
 \caption{ \bpiz: Projection of the fit result onto the m(\piz) (left) and \mmiss (right) variables.
\label{fig:datapi0fit}}
 \end{centering}
\end{figure}

\begin{figure}
 \begin{centering}
 \epsfig{file=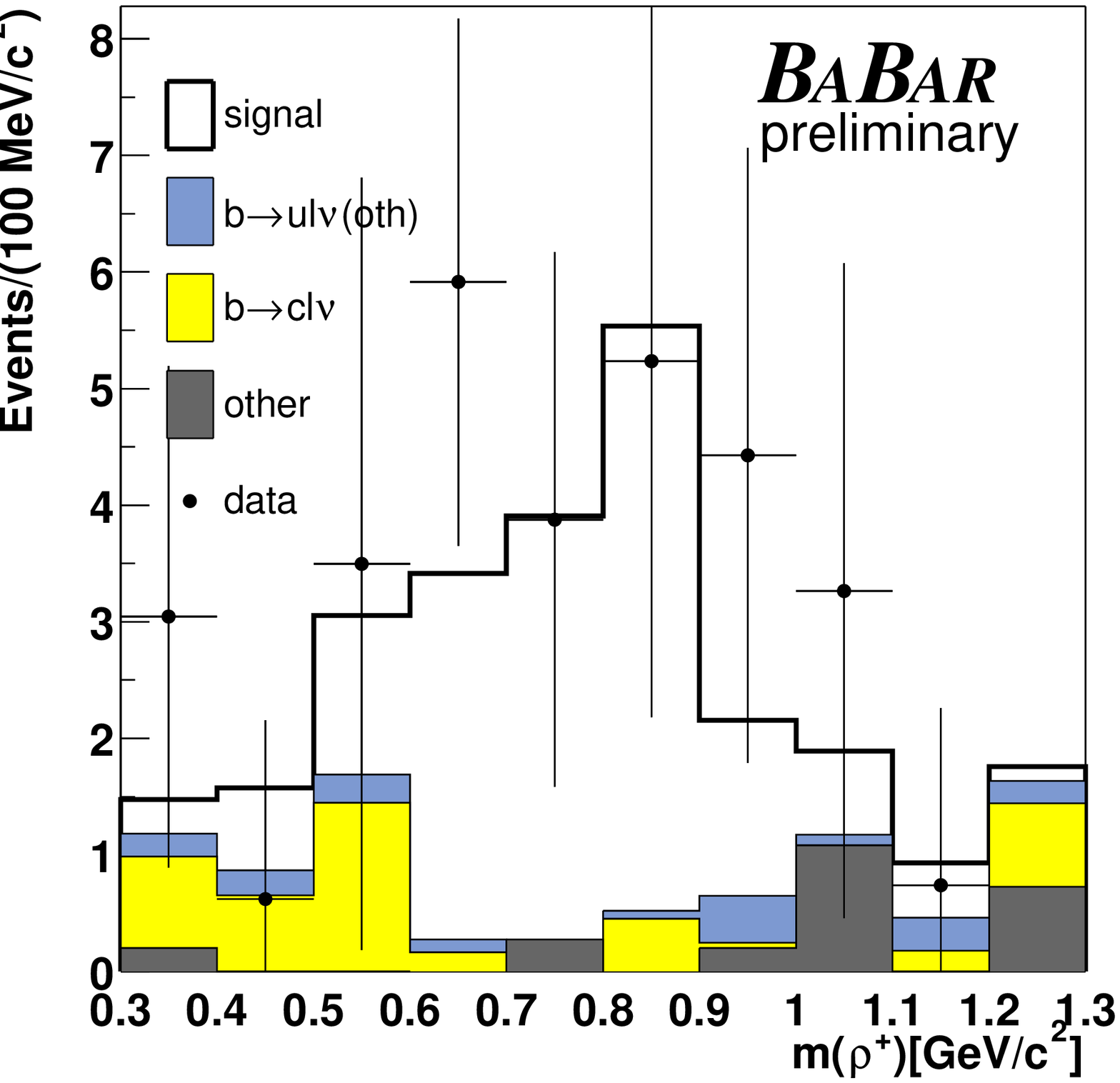,width=8.cm}        
 \epsfig{file=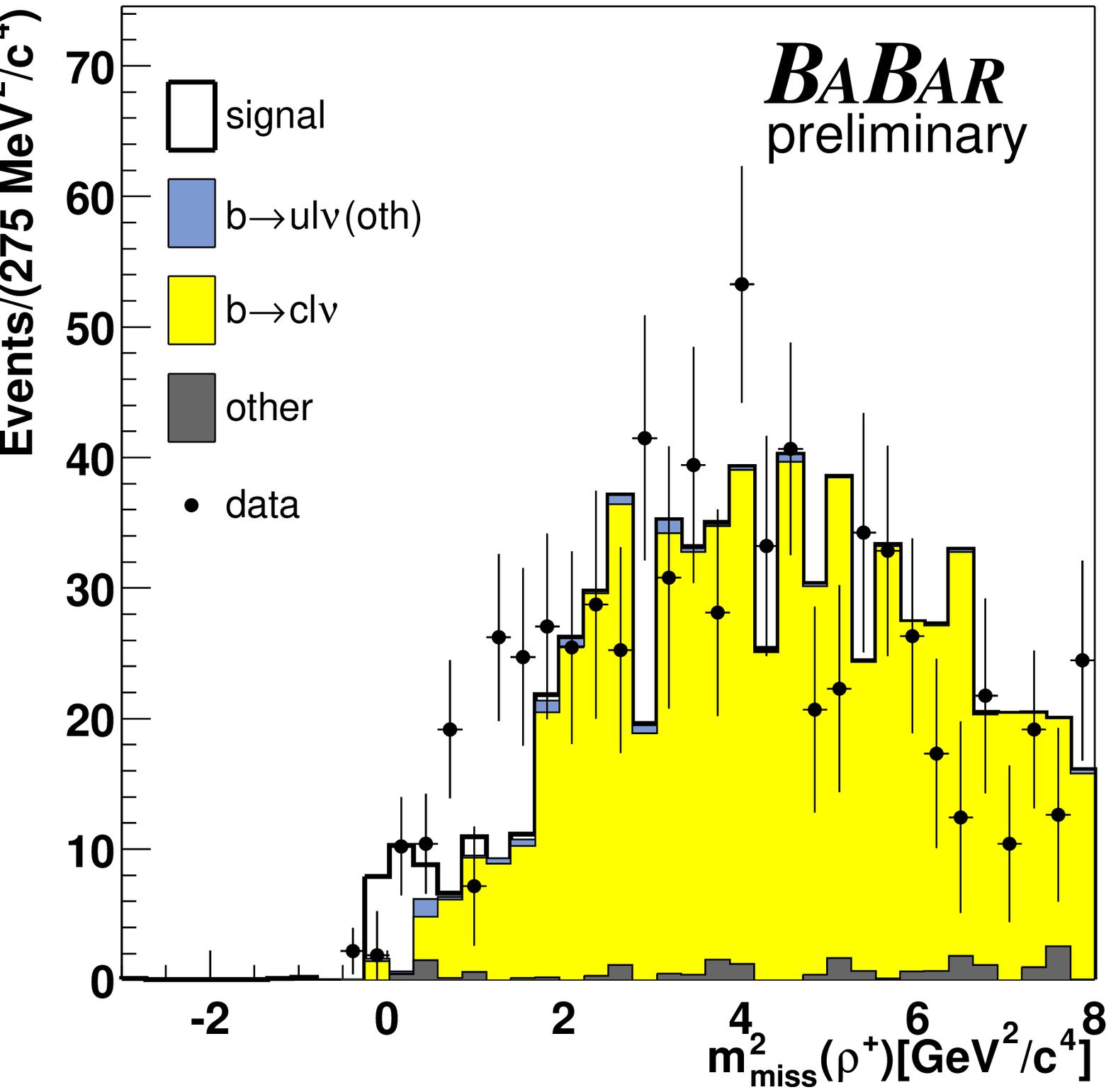,width=8.cm}         
 \caption{ \brhop: Projection of the fit result onto the m(\rhop) (left) and \mmiss (right) variables.
\label{fig:datarhofit}}
 \end{centering}
\end{figure}

\begin{figure}
 \begin{centering}
 \epsfig{file=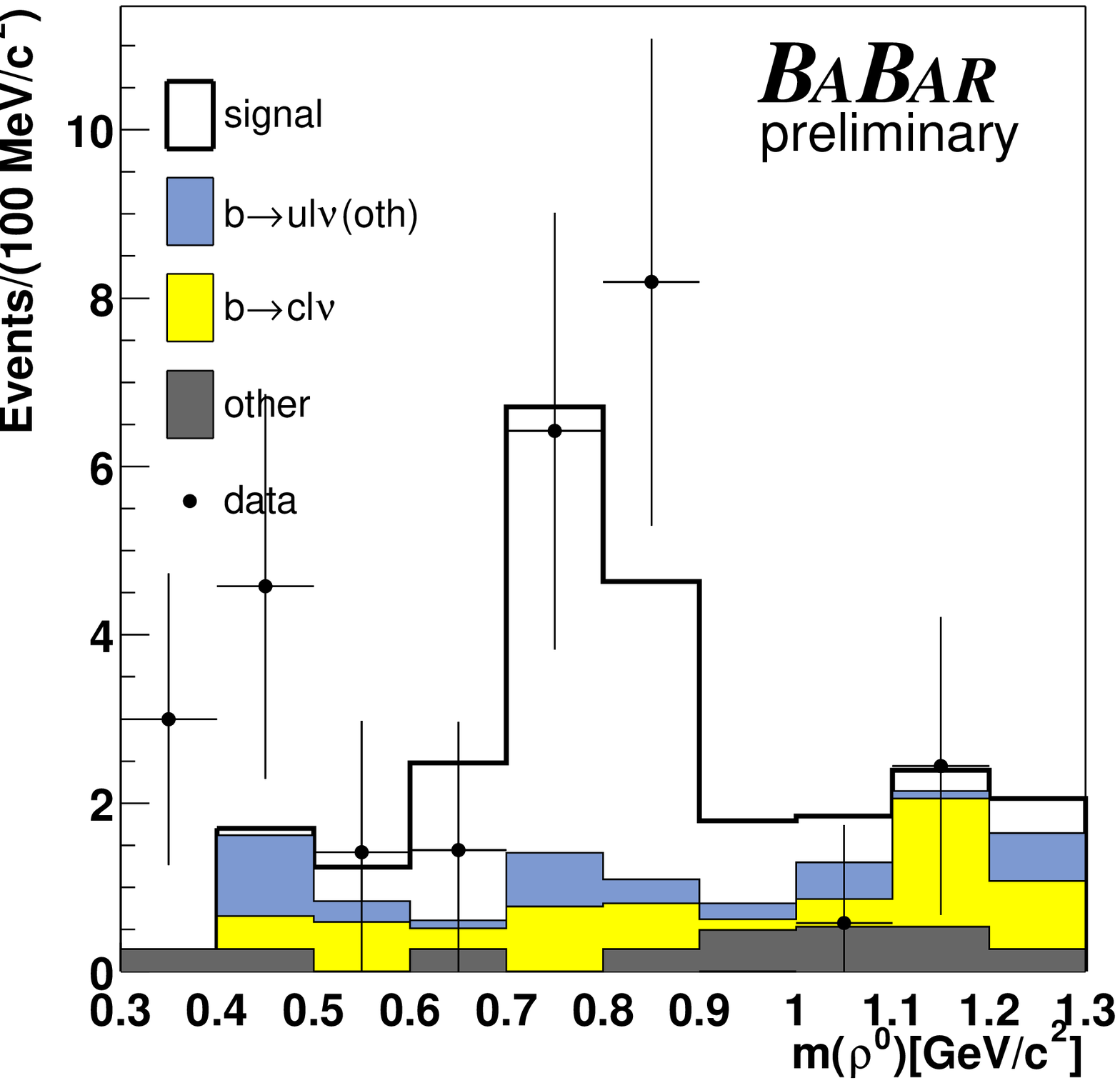,width=8.cm}       
 \epsfig{file=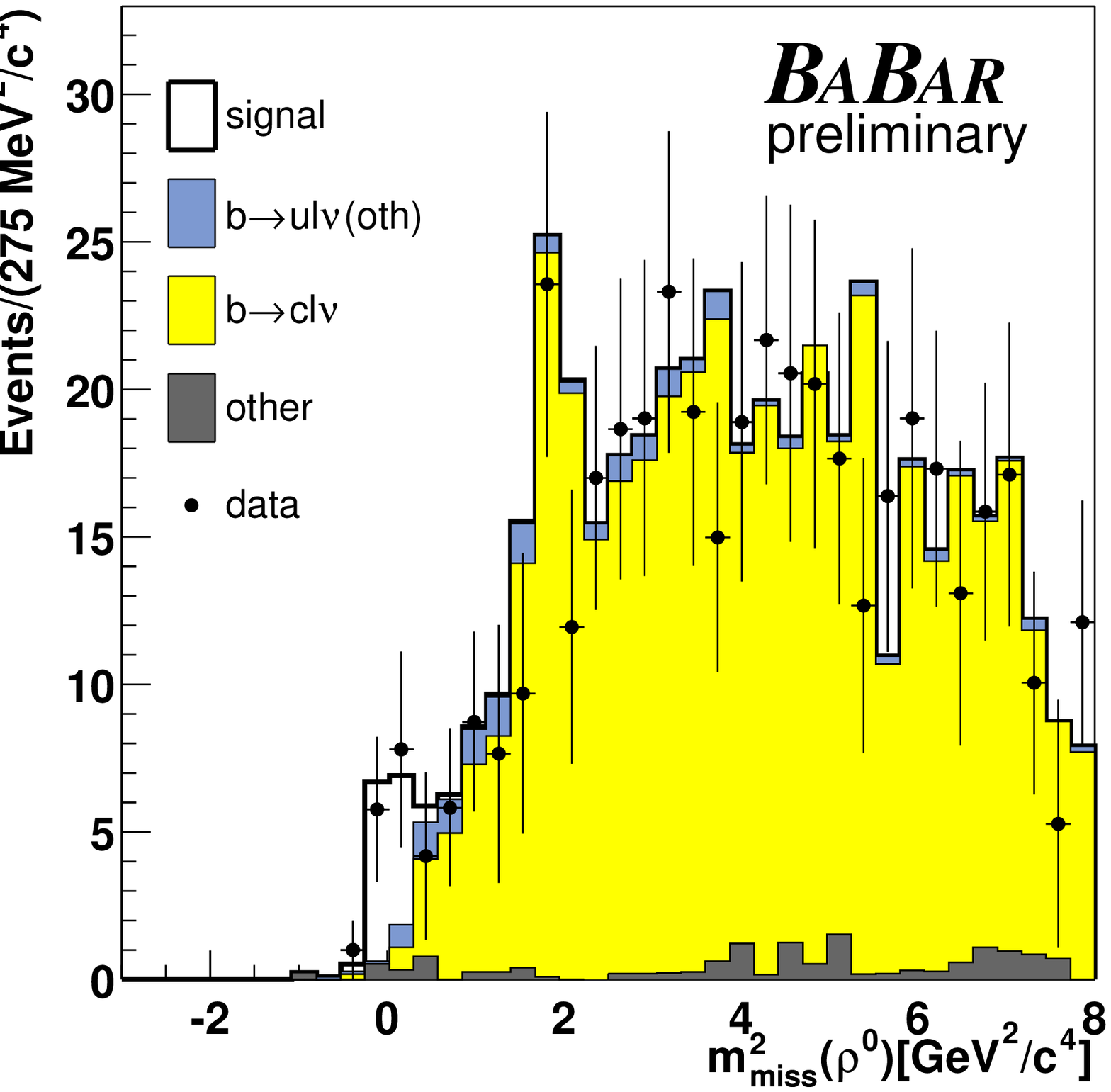,width=8.cm}        
 \caption{\brhozv: Projection of the fit result onto the m($pi^+\pi^-$) (left) and \mmiss (right) variables. Here
\rhozv means that no $B^-\rightarrow\pi^+\pi^- \ell \bar{\nu}$ non-resonant contribution is taken into 
account.
\label{fig:datarho0fit}}
 \end{centering}
\end{figure}

\begin{figure}
 \begin{centering}
 \epsfig{file=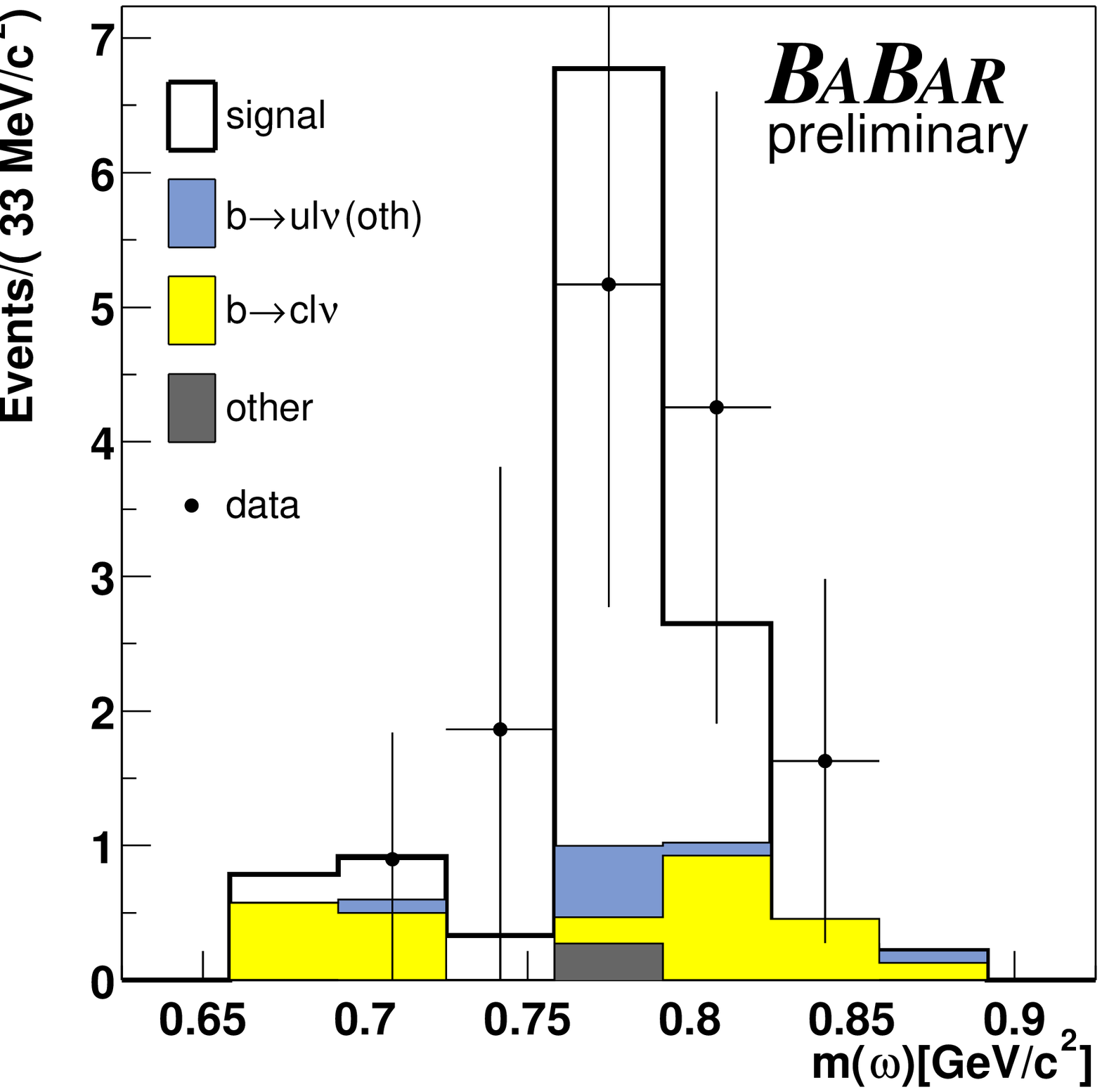,width=8.cm}      
 \epsfig{file=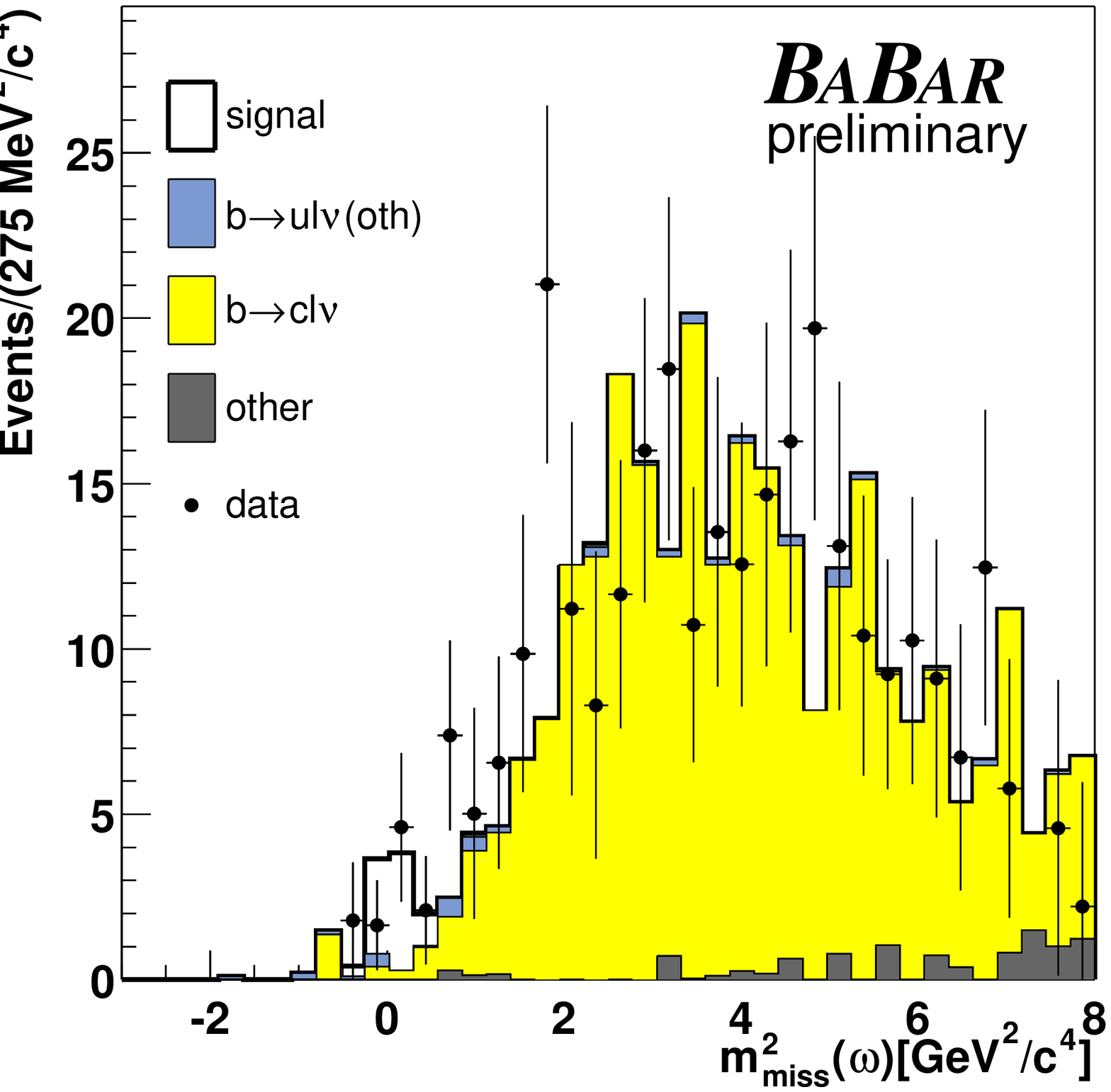,width=8.cm}       
 \caption{ \bomega: Projection of the fit result onto the m($\omega$) (left) and \mmiss (right) variables.
\label{fig:dataomegafit}}
 \end{centering}
\end{figure}

\begin{figure}
 \begin{centering}
 \epsfig{file=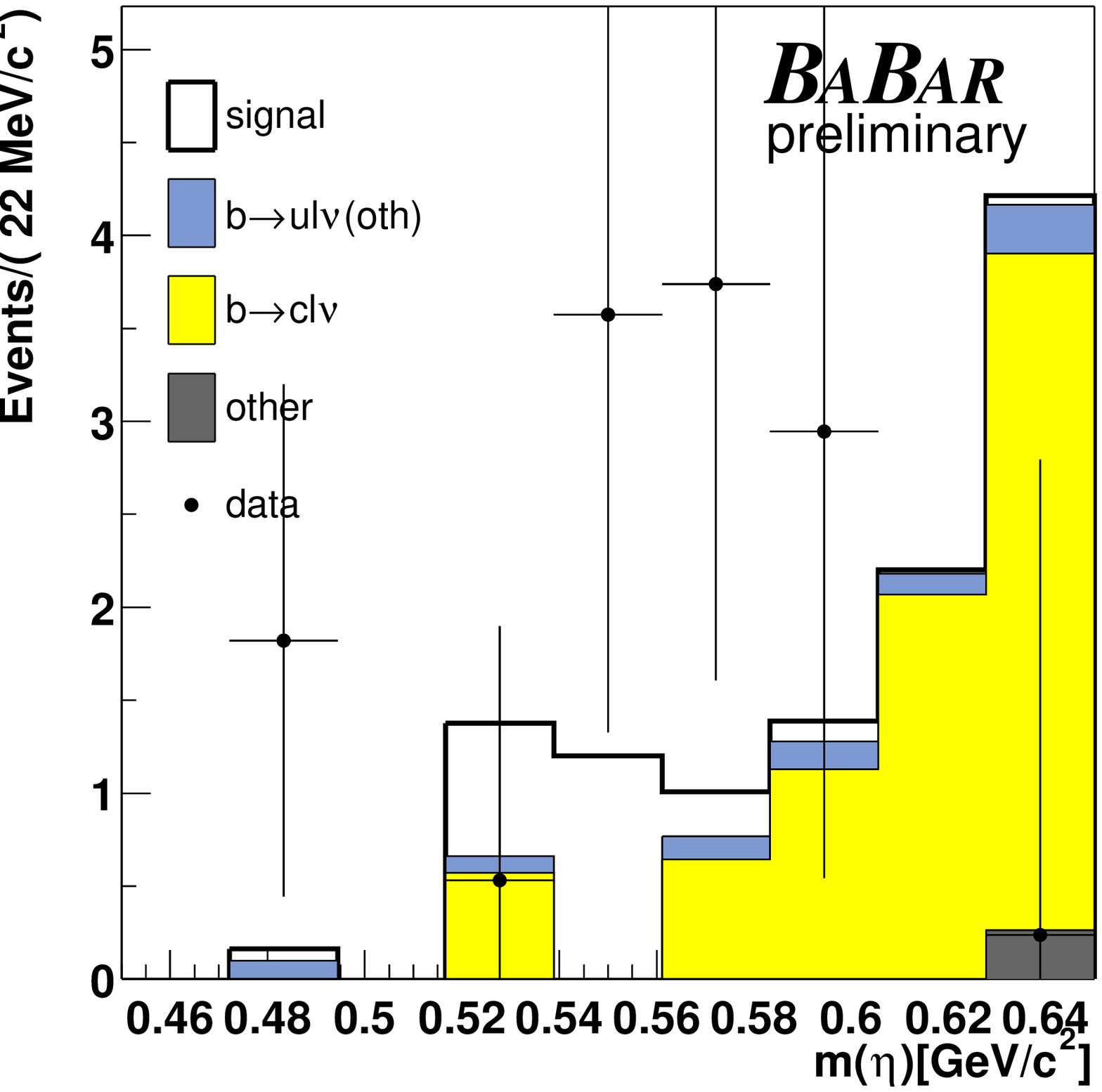,width=8.cm}        
 \epsfig{file=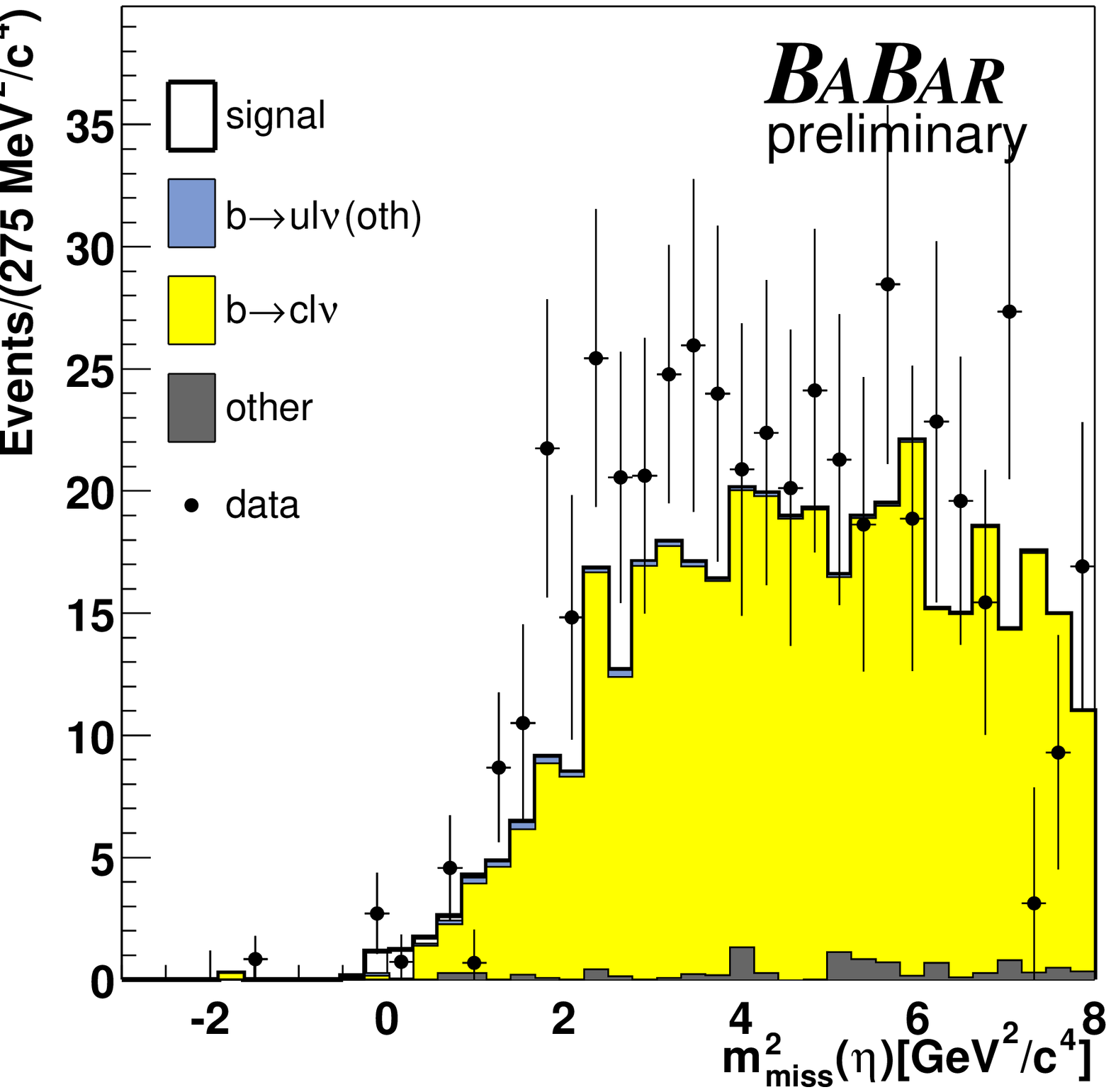,width=8.cm}         
 \caption{ \bet: Projection of the fit result onto the m($\eta$) (left) and \mmiss (right) variables.
\label{fig:dataetafit}}
 \end{centering}
\end{figure}

\begin{figure}
 \begin{centering}
 \epsfig{file=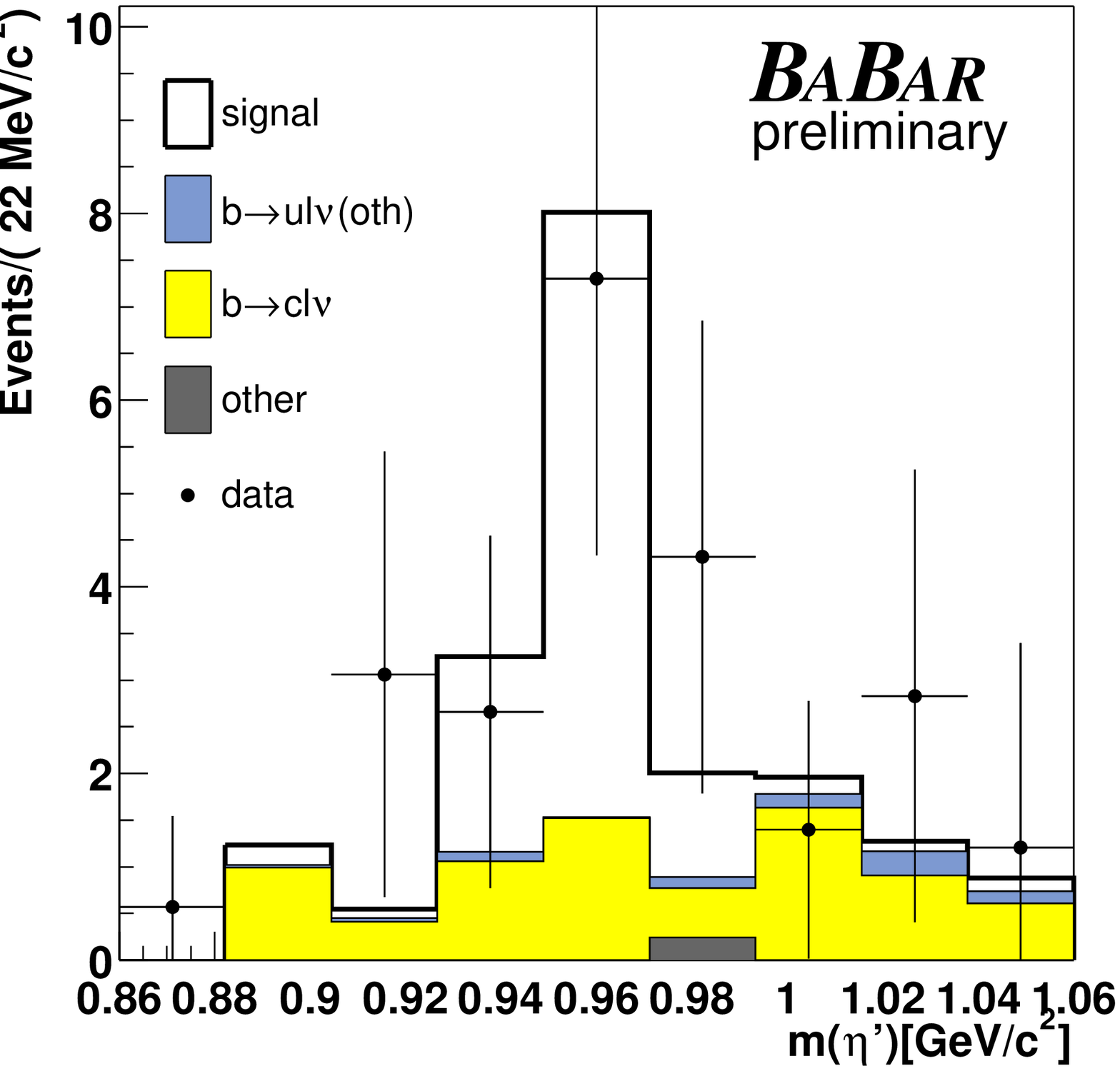,width=8.cm}       
 \epsfig{file=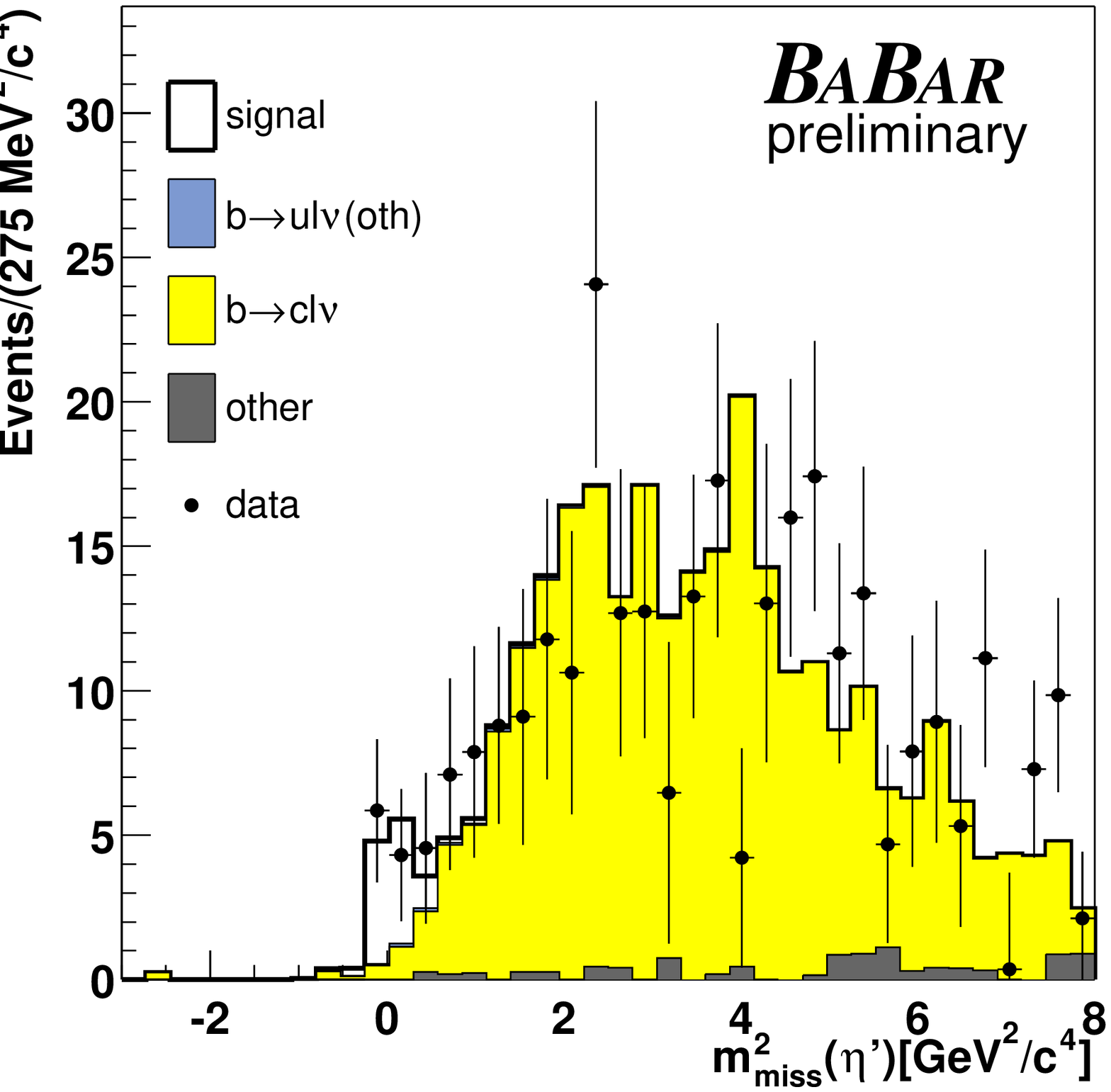,width=8.cm}        
 \caption{ \betp: Projection of the fit result onto the m(\etapr) (left) and \mmiss (right) variables.
\label{fig:dataetapfit}}
 \end{centering}
\end{figure}

\begin{figure}
 \begin{centering}
 \epsfig{file=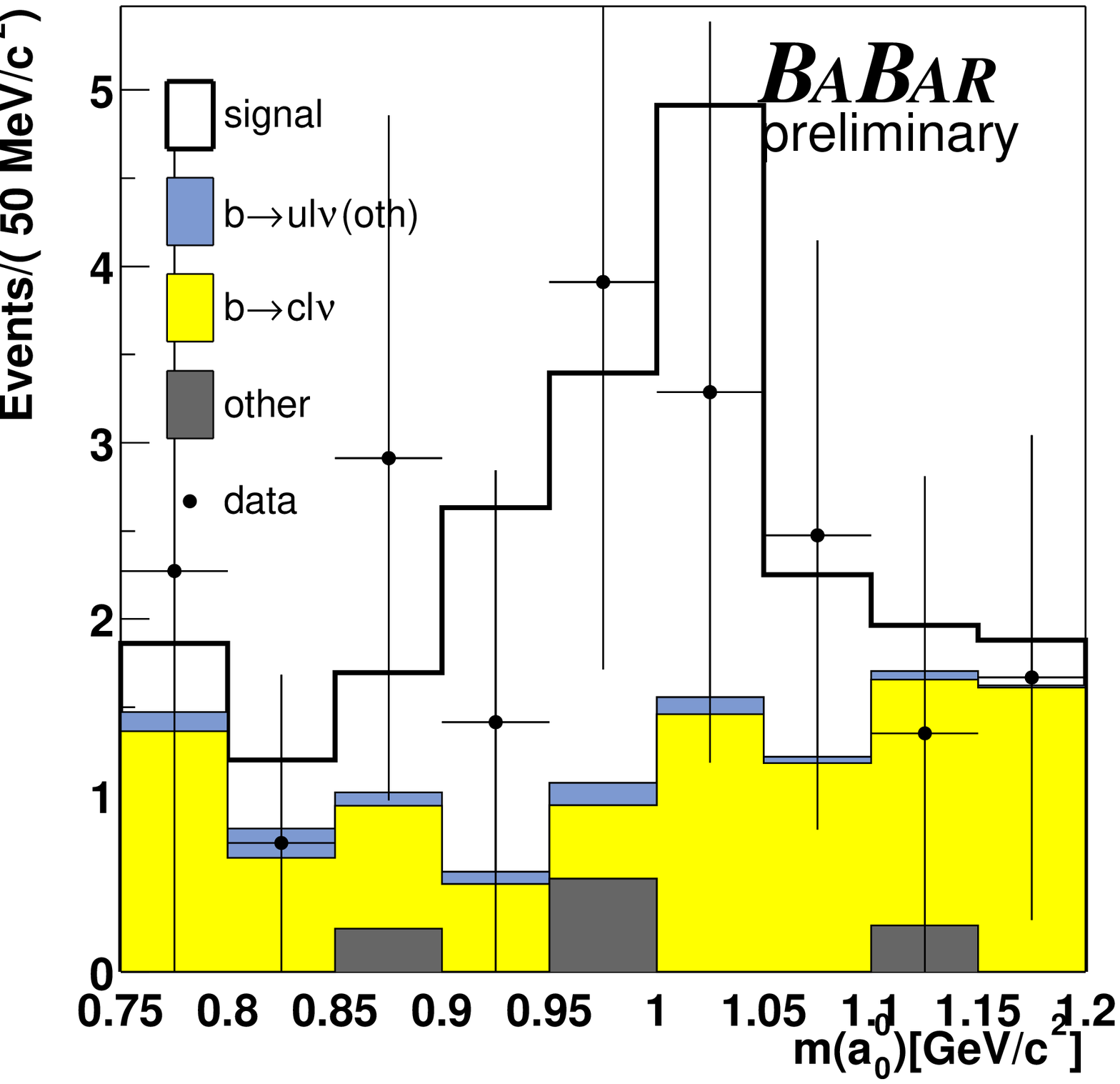,width=8.cm}         
 \epsfig{file=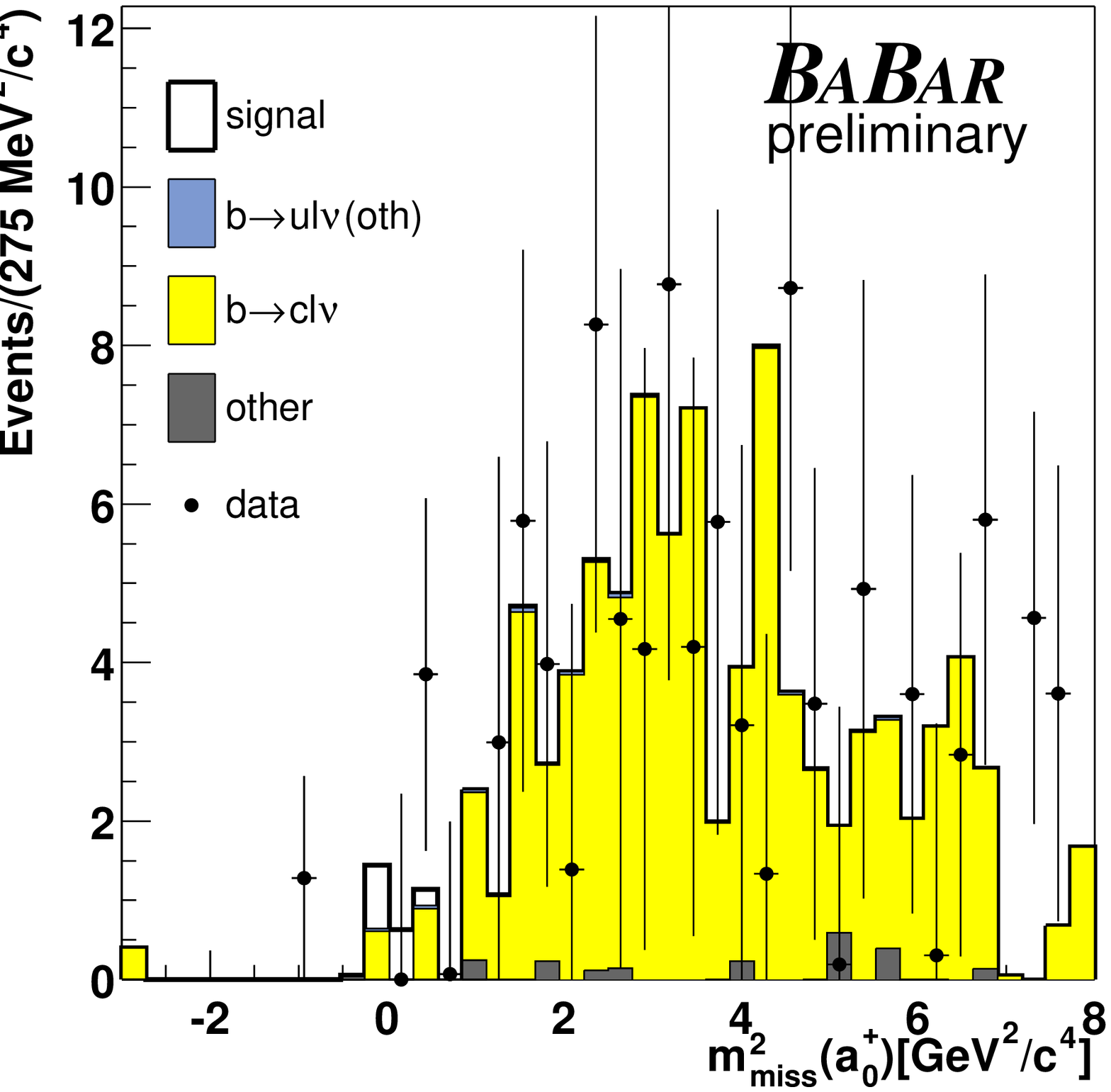,width=8.cm}  
 \caption{ \baz: Projection of the fit result onto the m(\az) (left) and \mmiss (right) variables.
\label{fig:dataa0fit}}
 \end{centering}
\end{figure}

\begin{figure}
 \begin{centering}
 \epsfig{file=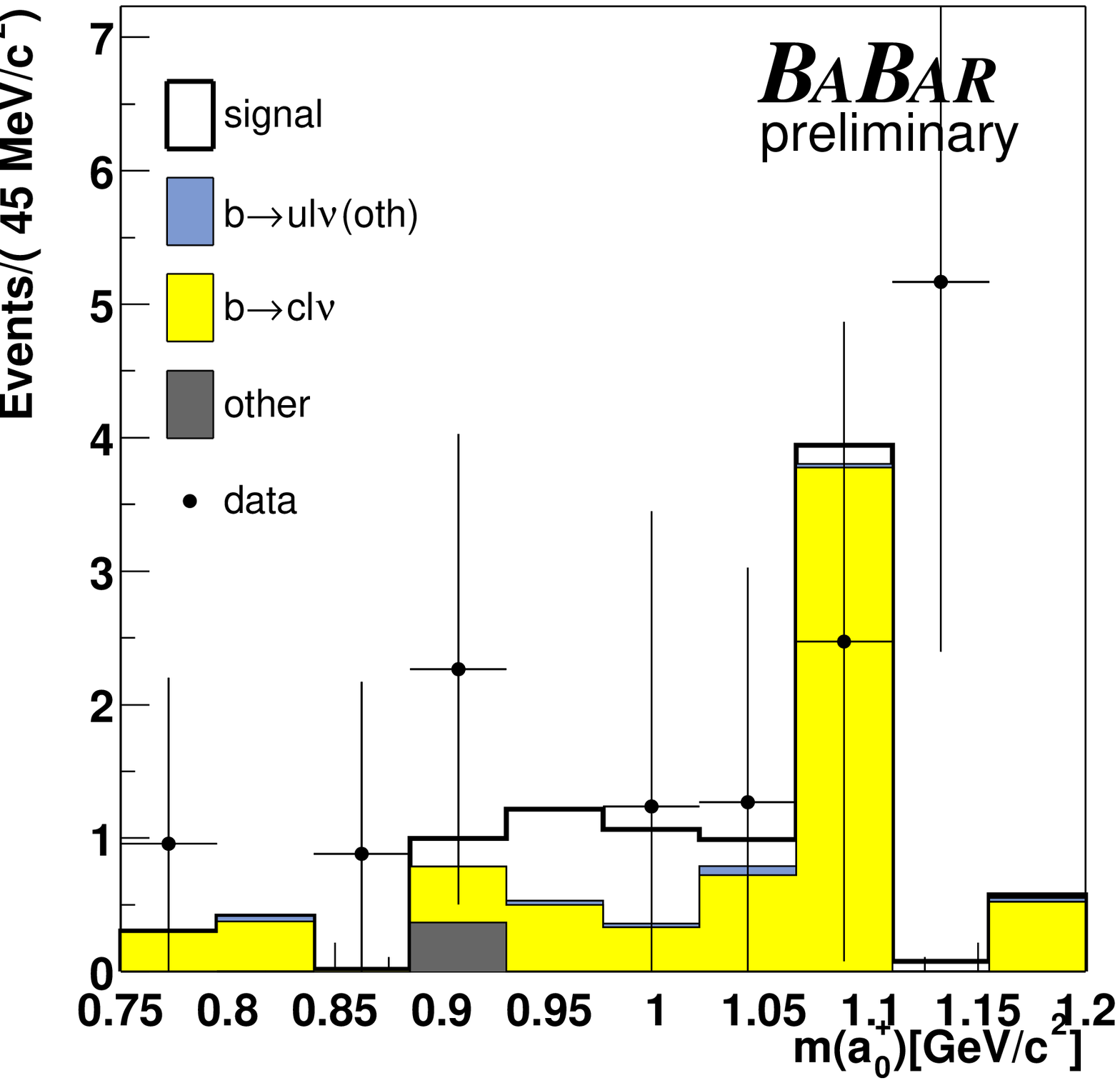,width=8.cm}        
 \epsfig{file=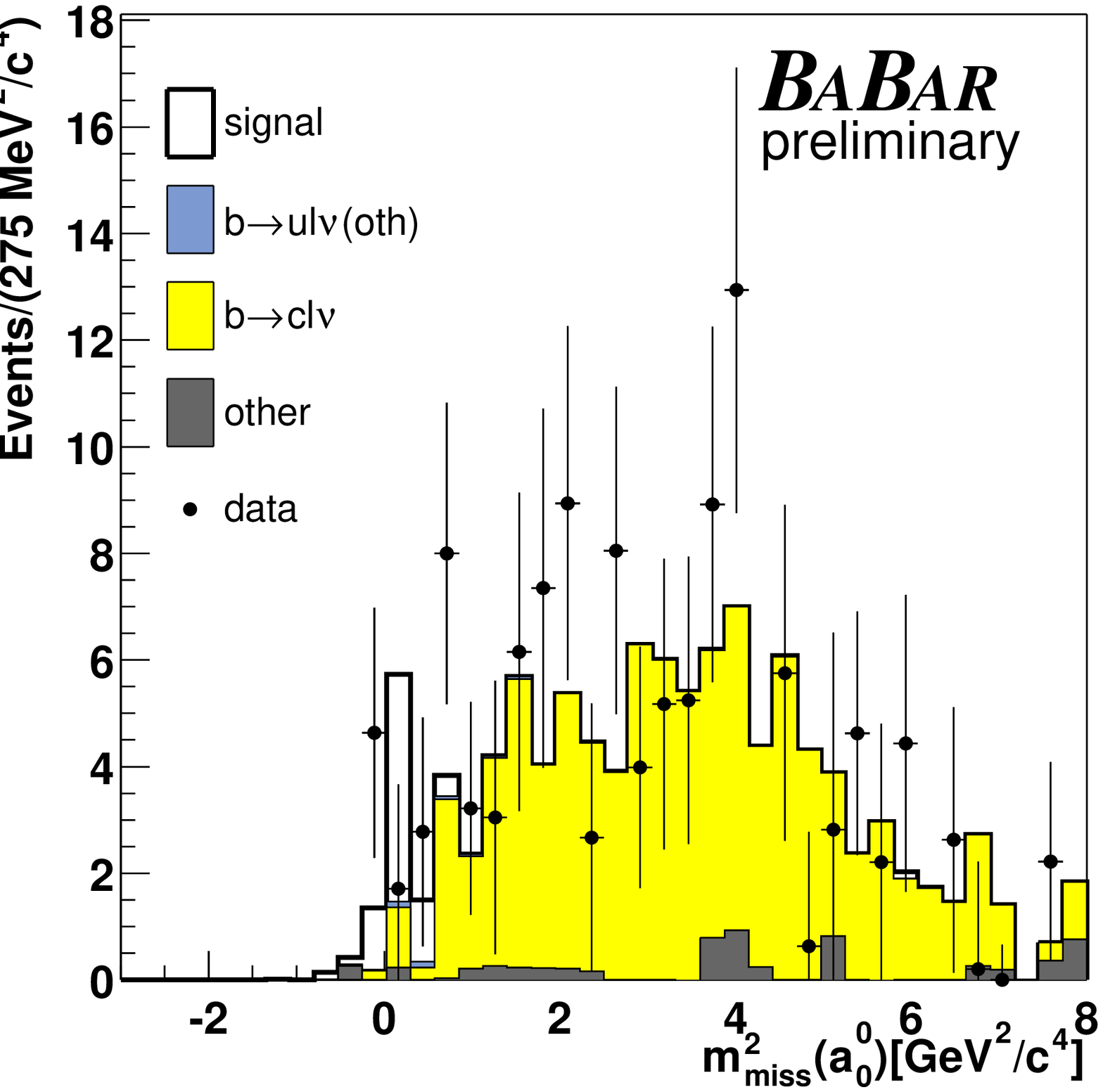,width=8.cm}         
 \caption{ \bazp: Projection of the fit result onto the m(\azp) (left) and \mmiss (right) variables.
\label{fig:dataa0pfit}}
 \end{centering}
\end{figure}

For the extraction of the ratio of branching ratios we use Equation~\ref{eq:ratioBR}. 
The only difference is in the inclusive semileptonic branching ratios since we use
$\BR(\Bnxlnu) = (10.4 \pm 0.3)\%$  and
$\BR(\Bpxlnu) = (11.3 \pm 0.3)\%$, obtained from the inclusive
semileptonic branching ratio~\cite{Aubert:2004aw} and the lifetime ratio
between neutral and charged $B$ mesons~\cite{Hagiwara:2002fs}.
The results are summarized in Table \ref{tab:fitsummary}.
In Figs.~\ref{fig:datapifit}-\ref{fig:dataa0pfit} the projections of the result on the reconstructed mass 
and on the \mmiss\ variable are shown for each mode. All
selection criteria are applied except for the ones on the variable that is
plotted.

\begin{table}[!hp]
\begin{center}
\caption{Fit results for all modes.}
\vspace{0.1in}
\begin{tabular}{|l|c|c|c|c|c|c|} 
\hline
           & $N_{excl}^{meas}$  & $BG_{excl}$  & $\epsilon_{sel}^{excl}$ & $N_{sl}^{meas}-BG_{sl}$ & $\frac{\epsilon_l^{sl} } {\epsilon_l^{excl}}$ & $R_{u/sl}[\times 10^{-3}]$  \\
\hline\hline
\bpi        & $ 11.1\pm3.9 $ & $ 0.9\pm0.5  $ & $ 0.65\pm0.03 $ & $ 15350\pm200 $ & $0.84\pm 0.01$ & $0.86\pm0.33 (\rm stat)$ \\
\bpiz       & $ 15.5\pm4.1 $ & $ 2.3\pm1.2  $ & $ 0.56\pm0.03 $ & $ 25250\pm300 $ & $0.87\pm 0.01$ & $0.81\pm0.25 (\rm stat)$ \\
\brhop      & $ 20.1\pm5.7 $ & $ 2.4\pm1.7  $ & $ 0.31\pm0.02 $ & $ 13100\pm200 $ & $0.76\pm 0.01$ & $3.3 \pm 1.1 (\rm stat)$ \\
\brhozv     & $ 15.7\pm4.3 $ & $ 4.0\pm1.0  $ & $ 0.45\pm0.03 $ & $ 22500\pm200 $ & $0.79\pm 0.01$ & $0.92\pm0.34 (\rm stat)$ \\
\bomega     & $ 9.3\pm3.3  $ & $ 1.7\pm0.8  $ & $ 0.21\pm0.02 $ & $ 25250\pm300 $ & $0.77\pm 0.01$ & $1.12\pm0.49 (\rm stat)$ \\
\bet        & $ 3.8\pm2.7  $ & $ 1.3\pm0.6  $ & $ 0.28\pm0.01 $ & $ 23050\pm200 $ & $0.87\pm 0.02$ & $0.34\pm0.36 (\rm stat)$ \\
\betp       & $ 13.9\pm4.2 $ & $ 4.3\pm1.2  $ & $ 0.14\pm0.01 $ & $ 23050\pm200 $ & $0.87\pm 0.02$ & $2.4 \pm 1.0 (\rm stat)$ \\
\baz        & $ 9.1\pm3.5  $ & $ 2.5\pm0.9  $ & $ 0.11\pm0.01 $ & $ 23050\pm200 $ & $0.91\pm 0.03$ & $2.4 \pm 1.3 (\rm stat)$ \\
\bazp       & $ 3.0\pm3.5  $ & $ 1.4\pm0.9  $ & $ 0.16\pm0.02 $ & $ 14700\pm200 $ & $0.87\pm 0.03$ & $0.6 \pm 1.4 (\rm stat)$ \\
\hline

\end{tabular}
\label{tab:fitsummary}
\end{center}
\end{table}

\clearpage

Since the branching ratios of the \az, \azp decays are not known,
the results for these channels will include the $\az \rightarrow \eta \piz$
and $\azp \rightarrow \eta \pip$ branching ratios. Note nevertheless
that they are known to be close to unity because they are dominant decay modes.

We obtain:
\begin{eqnarray}
 \BR(\bpi)&=&
(0.89\pm 0.34 (\rm stat.) \pm 0.12 (\rm sys.))\times 10^{-4}, \\ \nonumber
 \BR(\bpiz)&=&
(0.91 \pm 0.28 (\rm stat.) \pm 0.14 (\rm sys.))\times 10^{-4},\\ \nonumber
 \BR(\brhop)&=&
(3.5 \pm 1.1 (\rm stat.) \pm 0.7 (\rm sys.))\times 10^{-4},\\ \nonumber
 \BR(\brhozv)&=&
(1.04 \pm 0.39 (\rm stat.) \pm 0.16 (\rm sys.))\times 10^{-4},\\ \nonumber
 \BR(\bomega)&=&
(1.26 \pm 0.55 (\rm stat.) \pm 0.24 (\rm sys.))\times 10^{-4},\\ \nonumber
 \BR(\bet)&=&
(0.39 \pm 0.41 (\rm stat.) \pm 0.22 (\rm sys.)\times) 10^{-4},\\ \nonumber
 \BR(\betp)&=&
(2.7 \pm 1.2 (\rm stat.) \pm 0.5 (\rm sys.))\times 10^{-4},\\ \nonumber
 \BR(\baz) \BR(\az \rightarrow \eta \piz)&=&
(2.7 \pm 1.4 (\rm stat.) \pm 0.9 (\rm sys.))\times 10^{-4},\\ \nonumber
 \BR(\bazp) \BR(\azp \rightarrow \eta \pip)&=&
(0.7 \pm 1.6 (\rm stat.) \pm 0.3 (\rm sys.))\times 10^{-4},
\end{eqnarray}

\noindent where the errors on $\BR(\Bxlnu)$ are added in
quadrature. \\
The rates for exclusive decays involving \pip and \piz and \rhop, \rhoz, and $\omega$ can be constrained 
by the following isospin and quark model relations:
\begin{eqnarray}
\Gamma(\Bbar\rightarrow \pi\ell\nub) \equiv \Gamma(\bpi) &=& 2  \Gamma(\bpiz) \nonumber \\
\Gamma(\Bbar\rightarrow\rho\ell\nub) \equiv \Gamma(\brhop) &=& 2 \Gamma(\brhoz) \nonumber \\
\Gamma(\bomega) &=& \Gamma(\brhoz)
\end{eqnarray}
and it is possible to combine the results. Using a lifetime ratio of 
$\frac{\tau_{B^+}}{\tau_{B^0}}=1.086 \pm 0.017$ we obtain:  

\begin{eqnarray}
 \BR(\Bbar\rightarrow \pi\ell\nub)&=&
(1.08 \pm 0.28 (\rm stat.) \pm 0.16 (\rm sys.)\times) 10^{-4},\\ \nonumber
 \BR(\Bbar\rightarrow\rho\ell\nub)&=&
(2.57 \pm 0.52 (\rm stat.) \pm 0.59 (\rm sys.))\times 10^{-4}.
\end{eqnarray}

\vspace{0.7cm}
We also set 90\% C.L.  upper limits on the $\bet$, $\betp$, $\baz$, and $\bazp$  branching fractions:

\begin{eqnarray}
 \BR(\bet) &<& 1.2 \times 10^{-4}\ (90\%\ C.L.), \\ \nonumber
 \BR(\betp) &<& 4.5 \times 10^{-4}\ (90\%\ C.L.),\\ \nonumber
 \BR(\baz)\BR(\az \rightarrow \eta \piz)&<& 5.3 \times 10^{-4}\ (90\%\ C.L.),\\ \nonumber
 \BR(\bazp)\BR(\azp \rightarrow \eta \pip) &<& 3.3 \times 10^{-4}\ (90\%\ C.L.).
\end{eqnarray}

\section{Conclusions}
\label{sec:conclusions}
Preliminary results on charmless semileptonic $B$ decays have been obtained on a sample 
of 88 million \FourS\to\BB decays collected by the \babar\ experiment at the PEP-II asymmetric-energy $B$
factory at  SLAC, in events in which one $B$ meson decaying to a hadronic 
final state is fully reconstructed and the semileptonic decay of the second $B$ meson is identified 
by the detection of a charged lepton. 

We have explored several new approaches to the study of \Bxulnu\ decays, most of which are statistically limited and some of which
are already competitive with the best existing measurements.

From the measurement of the spectrum of the invariant mass of the hadrons $X_u$ (\mx) we derive the
branching fraction 
\begin{equation}
\BR(\Bxulnu)=(2.53\pm 0.29(\rm stat.)\pm 0.26(\rm
  sys.)^{+0.69}_{-0.41} (\rm theo.))\times 10^{-3}. 
\end{equation}
From the two-dimensional distribution of \mx\ and \Q, the squared invariant mass of the two leptons, 
we derive the partial branching fraction for $\mx<1.7\gevcc$ , $\Q>8\gevccsq$ to be 
\begin{equation}
\Delta\BR(\Bxulnu )=(0.88\pm 0.14(\rm stat.)\pm 0.13(\rm sys.)\pm 0.02(\rm
theo.))\times 10^{-3}.
\end{equation} 
From these two measurements, utilizing the calculation of the fraction of events in the selected phase space region
from DFN~\cite{DeFazio:1999sv} and  BLL~\cite{Bauer:2001yb} in the \mx\ and \mx-\Q\ analysis respectively,
 we can extract
\begin{eqnarray}
\label{eq:cleores}
\Vub _{\mx} & = & (4.77\pm 0.28 (\rm stat.)\pm 0.28 (\rm sys.)^{+0.69}_{-0.45} (\rm theo.))\times 10^{-3} ~~{\rm and}\\ \nonumber
\Vub  _{\mx-\Q}& = & (4.92\pm 0.39 (\rm stat.)\pm 0.36 (\rm sys.)\pm 0.46 (\rm theo.))\times 10^{-3}, 
\end{eqnarray}
respectively. These results are competitive with the current world average \footnote{
See the HFAG page http://www.slac.stanford.edu/xorg/hfag/semi/winter04/winter04.shtml . }$\Vub  =  (4.57\pm 0.61)\times 10^{-3}$.

The theoretical error is dominated by the uncertainty on the shape function parameters. In the process of writing this paper 
we were informed of an improved determination of these parameters from the photon energy spectrum measured by Belle in $b\to s\gamma$ decays 
(see Sec.~\ref{sec:theosys}).
Utilizing these new constraints on the shape function  parameters we get
\begin{eqnarray}
 \Vub _{\mx}&=&(5.22\pm 0.30(\rm stat.)\pm 0.31 (\rm sys.)^{+0.33}_{-0.32}(\rm theo.))\times 10^{-3}\\ \nonumber
\Vub _{\mx-\Q}& = & (4.98 \pm 0.40(\rm stat.) \pm 0.39 (\rm sys.)\pm 0.47(\rm theo.)) \times 10^{-3}, 
\end{eqnarray}
to be compared with the measured values of \Vub\ in Eq.~\ref{eq:cleores}. 
These results show that the determination of the shape function parameters
is critical to these measurements. Furthermore the  two approaches deal differently with the shape function parameterizations: in
particular
 the BLL approach does not correct the acceptance for shape function effects but assigns a large error.
The corresponding result is therefore
more stable but returns a worse error than the  DFN approach in case of accurate determinations of the shape function parameters.

On the other side,
the dependence of the \Vub\ measurement as a function of the requirement on \Q\  (Fig.~\ref{fig:mxq2dbr}) shows 
that both models reproduce the signal \Q\ distribution  well.

 The reconstructed \mx\ spectrum is also used to unfold the \mx\ distribution for 
\Bxulnu\ events and to measure its first and second moments (see Table~\ref{moments}).
With a higher statistics data sample, these measurements can be used to constrain the shape function parameters.

From the same data sample, several exclusive charmless semileptonic $B$ decays are identified 
and their branching fractions  measured (see Sec.~\ref{sec:exclresults}).
Imposing isospin and quark-model relationships we derive:
\begin{eqnarray}
\begin{array}{rclc}
\BR(\Bbar\rightarrow \pi\ell\nub)&=&
(1.08 \pm 0.28(\rm stat.) \pm 0.16(\rm sys.))\times 10^{-4}&,\\ \nonumber
\BR(\Bbar\rightarrow\rho\ell\nub)&=&
(2.57 \pm 0.52(\rm stat.) \pm 0.59(\rm sys.))\times 10^{-4}&.\\
\end{array}
\end{eqnarray}
Although these measurements are still statistically limited, the signal samples are very clean and are selected 
with very loose criteria which will allow the measurement
 of form factors with somewhat larger data samples.

We are grateful for the 
extraordinary contributions of our \pep2\ colleagues in
achieving the excellent luminosity and machine conditions
that have made this work possible.
The success of this project also relies critically on the 
expertise and dedication of the computing organizations that 
support \babar.
The collaborating institutions wish to thank 
SLAC for its support and the kind hospitality extended to them. 
This work is supported by the
US Department of Energy
and National Science Foundation, the
Natural Sciences and Engineering Research Council (Canada),
Institute of High Energy Physics (China), the
Commissariat \`a l'Energie Atomique and
Institut National de Physique Nucl\'eaire et de Physique des Particules
(France), the
Bundesministerium f\"ur Bildung und Forschung and
Deutsche Forschungsgemeinschaft
(Germany), the
Istituto Nazionale di Fisica Nucleare (Italy),
the Foundation for Fundamental Research on Matter (The Netherlands),
the Research Council of Norway, the
Ministry of Science and Technology of the Russian Federation, and the
Particle Physics and Astronomy Research Council (United Kingdom). 
Individuals have received support from 
CONACyT (Mexico),
the A. P. Sloan Foundation, 
the Research Corporation,
and the Alexander von Humboldt Foundation.

\begin{appendix}
%
\section{The Unfolded Spectrum and the Covariance Matrix}
\label{app_tables}
%

\begin{table}[h]
\centering
\renewcommand{\arraystretch}{1.15}
\caption{Fraction of \Bxulnu\ events measured in a given bin $i$ of Fig.~\ref{unf:result} and corresponding breakdown of errors.
All numbers are in percent.
\label{tab:spectrum}}
\begin{tabular}{rcccccl}
\hline\hline
$i$ & $x_\mathrm{unf}^i$ & $\sigma^{i}$ & $\sigma_{\mathrm{stat}}^i\oplus  \sigma^i_\upr{MCstat}$ &
$\sigma_{\mathrm{det}}^i$ & $\sigma^i_{\mathrm{ul\nu}}\oplus \sigma^i_\upr{theo}$ &
$\sigma^i_\upr{bkg} \oplus \sigma^i_\upr{breco}$\\
\hline
1 & 2.890 & 1.655 & 1.206 & 0.589 & 0.322 & 0.423\\
2 & 1.067 & 1.294 & 0.895 & 0.415 & 0.685 & 0.266\\
3 & 5.355 & 4.039 & 3.325 & 0.906 & 0.952 & 1.026\\
4 & 9.840 & 4.934 & 3.532 & 1.388 & 1.999 & 1.537\\
5 & 28.024 & 8.330 & 5.302 & 2.807 & 3.111 & 4.183\\
6 & 25.830 & 3.657 & 1.850 & 1.223 & 1.169 & 2.377\\
7 & 15.880 & 5.996 & 3.964 & 1.806 & 2.742 & 1.872\\
8 & 7.189 & 5.742 & 3.935 & 1.940 & 1.161 & 2.998\\
9 & 2.742 & 4.213 & 2.534 & 1.276 & 1.939 & 2.209\\
10 & 0.880 & 1.822 & 1.194 & 0.627 & 0.290 & 1.097\\
11 & 0.261 & 0.737 & 0.478 & 0.260 & 0.104 & 0.451\\
12 & 0.038 & 0.149 & 0.088 & 0.064 & 0.036 & 0.084\\
13 & 0.002 & 0.018 & 0.006 & 0.011 & 0.008 & 0.006\\
14 & 0.001 & 0.005 & 0.002 & 0.003 & 0.002 & 0.002\\
15 & 0.000 & 0.001 & 0.000 & 0.000 & 0.000 & 0.000\\
16 & 0.000 & 0.000 & 0.000 & 0.000 & 0.000 & 0.000\\
\hline\hline
\end{tabular}
\end{table}

\begin{landscape}
\begin{table}
\centering
\caption{Full covariance matrix on the fraction of \Bxulnu\ in the  unfolded \mx spectrum (as in Table~\ref{tab:spectrum})). All
numbers are  in $10^{-4}$. A version with a higher number of digits can be obtained from the authors.}
\renewcommand{\arraystretch}{1.15}
\begin{tabular}{cccccccccccccccc}
\hline \hline 
2.739&1.021&4.254&5.685&8.257&-1.605&-7.474&-6.267&-4.099&-1.734&-0.647&-0.117&-0.010&-0.003&-0.000&-0.000\\
1.021&1.674&2.975&3.853&4.058&-0.992&-4.398&-4.012&-2.465&-1.153&-0.460&-0.091&-0.009&-0.003&-0.000&-0.000\\
4.254&2.975&16.314&13.999&17.275&-5.302&-17.648&-15.958&-9.264&-4.448&-1.802&-0.351&-0.033&-0.009&-0.002&-0.001\\
5.685&3.853&13.999&24.343&34.115&-1.585&-28.030&-24.898&-17.030&-7.222&-2.692&-0.484&-0.039&-0.012&-0.002&-0.001\\
8.257&4.058&17.275&34.115&69.389&11.837&-45.085&-45.370&-33.773&-14.300&-5.325&-0.965&-0.082&-0.025&-0.004&-0.002\\
-1.605&-0.992&-5.302&-1.585&11.837&13.373&-0.222&-5.212&-6.439&-2.675&-0.980&-0.176&-0.015&-0.005&-0.001&-0.000\\
-7.474&-4.398&-17.648&-28.030&-45.085&-0.222&35.949&31.214&22.623&9.146&3.289&0.574&0.045&0.014&0.002&0.001\\
-6.267&-4.012&-15.958&-24.898&-45.370&-5.212&31.214&32.971&22.363&10.272&4.042&0.763&0.067&0.020&0.003&0.002\\
-4.099&-2.465&-9.264&-17.030&-33.773&-6.439&22.623&22.363&17.751&7.207&2.603&0.466&0.041&0.012&0.002&0.001\\
-1.734&-1.153&-4.448&-7.222&-14.300&-2.675&9.146&10.272&7.207&3.319&1.309&0.249&0.022&0.007&0.001&0.001\\
-0.647&-0.460&-1.802&-2.692&-5.325&-0.980&3.289&4.042&2.603&1.309&0.543&0.107&0.010&0.003&0.001&0.000\\
-0.117&-0.091&-0.351&-0.484&-0.965&-0.176&0.574&0.763&0.466&0.249&0.107&0.022&0.002&0.001&0.000&0.000\\
-0.010&-0.009&-0.033&-0.039&-0.082&-0.015&0.045&0.067&0.041&0.022&0.010&0.002&0.000&0.000&0.000&0.000\\
-0.003&-0.003&-0.009&-0.012&-0.025&-0.005&0.014&0.020&0.012&0.007&0.003&0.001&0.000&0.000&0.000&0.000\\
-0.000&-0.000&-0.002&-0.002&-0.004&-0.001&0.002&0.003&0.002&0.001&0.001&0.000&0.000&0.000&0.000&0.000\\
-0.000&-0.000&-0.001&-0.001&-0.002&-0.000&0.001&0.002&0.001&0.001&0.000&0.000&0.000&0.000&0.000&0.000\\
\hline \hline
\end{tabular}
\label{textcovmat}
\end{table}
\end{landscape}

\end{appendix}


\bibliographystyle{physrev4wt}

\begin{thebibliography}{10}

\bibitem{Aubert:2003zw}
\babar, B.~Aubert {\em et~al.},
\newblock Phys. Rev. Lett. {\bf 92}, 071802 (2004).

\bibitem{Bauer:2003pi}
C.~W. Bauer and A.~V. Manohar, ``Shape function effects in $B \to X_s \gamma$
  and $B \rightarrow X_u l \nu$ decays'',
\newblock hep-ph/0312109.

\bibitem{Bosch:2004bt}
S.~W. Bosch, B.~O. Lange, M.~Neubert and G.~Paz, ``Proposal for a precision
  measurement of $|V(ub)|$'',
\newblock hep-ph/0403223.

\bibitem{Gibbons:2004dg}
L.~Gibbons, ``The status of $|V(ub)|$'',
\newblock hep-ex/0402009.

\bibitem{DeFazio:1999sv}
F.~De~Fazio and M.~Neubert,
\newblock JHEP {\bf 06}, 017 (1999).

\bibitem{Bauer:2001yb}
C.~W. Bauer, Z.~Ligeti and M.~Luke, ``Precision determination of V(ub)'',
\newblock hep-ph/0111387.

\bibitem{Aubert:2001tu}
\babar, B.~Aubert {\em et~al.},
\newblock Nucl. Instrum. Meth. {\bf A479}, 1 (2002).

\bibitem{geant}
GEANT4, S.~Agostinelli {\em et~al.},
\newblock Nucl. Instrum. Meth. {\bf A506}, 250 (2003).

\bibitem{Cronin-Hennessy:2001fk}
CLEO, D.~Cronin-Hennessy {\em et~al.},
\newblock Phys. Rev. Lett. {\bf 87}, 251808 (2001).

\bibitem{Sjostrand:1994yb}
T.~Sjostrand, ``High-energy physics event generation with PYTHIA 5.7 and JETSET
  7.4'',
\newblock Comput. Phys. Commun. {\bf 82}, 74 (1994).

\bibitem{isgw2}
N.~Isgur and D.~Scora,
\newblock Phys. Rev. D {\bf 52}, 2783 (1995).

\bibitem{tadao}
A.~Limosani and T.~Nozaki, ``Extraction of the $b$-quark shape function
  parameters using the Belle $B\to X_s\gamma$ photon energy spectrum.'',
\newblock hep-ex/0407052.

\bibitem{Duboscq:1996mv}
CLEO, J.~E. Duboscq {\em et~al.},
\newblock Phys. Rev. Lett. {\bf 76}, 3898 (1996).

\bibitem{Goity:1995xn}
J.~L. Goity and W.~Roberts,
\newblock Phys. Rev. {\bf D51}, 3459 (1995).

\bibitem{Albrecht:1993pu}
ARGUS, H.~Albrecht {\em et~al.},
\newblock Phys. Lett. {\bf B318}, 397 (1993).

\bibitem{cry}
Crystal Ball, T.~Skwarnicki, ``A Study of the Radiative Cascade Transitions
  Between the $\Upsilon^\prime$ and $\Upsilon$ Resonances'',
\newblock DESY F31-86-02, Ph.D. thesys.

\bibitem{Aubert:2004aw}
BABAR, B.~Aubert {\em et~al.},
\newblock Phys. Rev. Lett {\bf 93}, 011803 (2004).

\bibitem{Hagiwara:2002fs}
Particle Data Group, K.~Hagiwara {\em et~al.},
\newblock Phys. Rev. {\bf D66}, 010001 (2002).

\bibitem{PDG04}
Particle Data Group, S.~Eidelman {\em et~al.},
\newblock {Physics Letters B} {\bf 592}, 1 (2004).

\bibitem{Hocker:1996kb}
A.~H{\oe}cker and V.~Kartvelishvili,
\newblock Nucl. Instrum. Meth. {\bf A372}, 469 (1996).

\bibitem{Teige:1996fi}
E852, S.~Teige {\em et~al.},
\newblock Phys. Rev. {\bf D59}, 012001 (1999).

\bibitem{Athar:2003yg}
CLEO, S.~B. Athar {\em et~al.},
\newblock Phys. Rev. {\bf D68}, 072003 (2003).

\end{thebibliography}

\cleardoublepage

\end{document}